\documentclass[preprint,12pt]{elsarticle}

\usepackage{amssymb}
\usepackage{amsmath}

\usepackage{latexsym}
\usepackage{amsthm}
\usepackage{booktabs}
\usepackage{enumitem}
\usepackage{graphicx}
\usepackage{color}
\usepackage{adjustbox}
\usepackage{comment}
\usepackage[numbers]{natbib}
\usepackage[english]{babel}

\usepackage{ifthen}  
\usepackage{algorithm}
\usepackage{algpseudocode}
\usepackage{adjustbox}

\algnewcommand\KwTo{\textbf{ to }}
\algnewcommand\KwOr{\textbf{ or }}
\algnewcommand\KwAnd{\textbf{ and }}
\algrenewcommand\alglinenumber[1]{\scriptsize #1:}

\newtheorem{theorem}{Theorem}

\newtheorem{proposition}[theorem]{Proposition}

\newtheorem{definition}{Definition}
\newtheorem{remark}{Remark}

\journal{Results in Engineering}

\begin{document}

\begin{frontmatter}



\title{A Framework for Reputation Aware Uninorm-driven Consensus Algorithms for Blockchain Networks}

\author[1]{Bruno Ramos-Cruz} 
\ead{brcruz@ujaen.es}

\author[1,2]{Javier Andreu-Perez}
\cortext[1]{Corresponding author: j.andreu-perez@essex.ac.uk (Javier Andreu-Perez) }
\ead{j.andreu-perez@essex.ac.uk}

\author[2]{David Richerby}
\ead{david.richerby@essex.ac.uk}

\author[1]{Luis Martínez}
\ead{martin@ujaen.es}

\affiliation[1]{organization={Computer Science Department, University of Jaen},
            addressline={Jaen}, 
            postcode={23071}, 
            country={Spain}}

\affiliation[2]{organization={School of Computer Science and Electronic Engineering, University of Essex},
    city={Colchester},
    postcode={CO4 3SQ}, 
    country={United Kingdom}}

\begin{abstract}
The operation of blockchain is governed by consensus algorithms (CA). Several consensus mechanisms require significant computational power, while others necessitate high amounts of stakes to select the participant to validate and verify the transactions in the block, leading to centralisation of power and participant exclusion. This paper proposes a novel methodology to address these issues in reputation-based consensus algorithms by studying  the reputation behaviour of the validator using intuitionistic fuzzy sets (IFSs)  and uninorm aggregation operations (UAOs). Our approach uses IFSs to express the ''reputation" because the reputation values in a consensus algorithm eventually imply uncertainty, and IFSs facilitate the representation of a lack of precise knowledge about reputation. Moreover, this methodology utilises uninorm aggregation operations to monitor reputation over time and reinforces the importance of negative and positive reputation. Consequently, this solution allows validators to rectify past failures in subsequent verification processes and foster an equitable consensus algorithm design. The proposed framework maintains linear computational complexity and does not introduce additional communication overhead beyond the underlying consensus protocol. Supported by experimental results, our methodology demonstrates improved performance and evaluation, promising advancements in blockchain network fairness and inclusivity. 
\end{abstract}




\end{frontmatter}



\section{Introduction}
Blockchain technology has significantly transformed diverse industries and institutions, offering decentralised and transparent systems for transactions and data management \cite{CHEN20221}. However, despite its potential, some challenges persist within blockchain networks, particularly in the realm of consensus algorithms \cite{Espinoza2022,Fahim2023}. These algorithms define rules for nodes on a distributed network, which ensures the validity and security of transactions, as well as maintaining trust and suitable network functioning. Consensus algorithms such as Proof of Work (PoW) \cite{Back2002, Nakamoto2008}, Proof of Stake (PoS) \cite{Ethereum2023}, and Delegated Proof of Stake (DPoS) \cite{Larimer2013}, among others, face hurdles related to the selection of nodes to carry out the block validation and verification process. For instance, PoW demands substantial computational power, PoS operates on a stake-based incentive model in which the nodes with the highest stake have a higher probability of being chosen, and DPoS allows network users vote for delegates to validate blocks, affecting decentralisation. Furthermore, for reputation-based consensus algorithms \cite{Ethereum2023, Larimer2013}, if a node loses its reputation, it can never again participate in the validation process. These challenges often lead to the concentration of power among a specific group of participants, consequently resulting in centralisation and the exclusion of smaller stakeholders, posing risks to the integrity and security of the network.

Considering the challenges, this contribution focusses on studying the behaviour of the nodes involved in the block validation and verification process. To do so, a novel methodology is presented to monitor the reputation behaviour of the nodes in the consensus algorithm, integrating Intuitionistic Fuzzy Sets (IFSs) and Uninorm Aggregation Operations (UAOs) into blockchain networks. 

When discussing reputation in a consensus algorithm, the reputation values eventually imply uncertainty, and this proposal advocates the use of IFSs because the knowledge about reputation in a blockchain network cannot be complete. There is a relative uncertainty of the information about it; for this reason, IFSs facilitate the representation of a lack of knowledge about a concept that, in this case, will be \textit{reputation}. Moreover, in consensus algorithms where the reputation fluctuates, increasing and decreasing \cite{Ramos2024}, classical algorithms typically reduce the reputation without providing a mechanism for recovery. In contrast, this methodology introduces a reputation recovery mechanism based on UAOs, reinforcing nodes that demonstrate positive behaviour while penalising those exhibiting negative behaviour within the network.

UAO are mathematical functions used in fuzzy logic to combine multiple input values into a single output value. These operators present very interesting properties and play a crucial role in synthesising information from various sources. They are particularly useful for handling uncertain or imprecise data. The proposed methodology uses UAO to reinforce both the evolution of positive and negative reputations over time, computing a reputation weight for each node in the consensus algorithm that takes into account current and previous results in the validation process. Additionally, using UAO, validators' reputation is monitored over time. This dynamic approach allows participants to rectify past mistakes, regain reputation, and promote a more equitable and inclusive consensus algorithm design. 

The \textbf{main novelties of this paper} are:

\begin{itemize}
\item An innovative methodology for reputation aware uninorm-driven consensus algorithms for blockchain.
\item An intuitionistic framework for managing reputation uncertainty.
\item A reputation recovery mechanism that reinforces both positive and negative reputation evolution over time.
\item An analysis to define the suitable functions used in both IFSs and UOAs for the consensus algorithm in blockchain.
\end{itemize}

Moreover, through empirical validation, an illustrative example is presented to show the performance of the proposed approach, showcasing its potential to offer a new paradigm for enhancing resilience, security, and diversification on the blockchain. 

The paper is structured as follows:  Section \ref{Sec:background} provides background on blockchain, intuitionistic fuzzy sets, and uninorm aggregation operations. Section \ref{Sec:rw} reviews related work. Section \ref{Sec:methodology} details the methodology. Section \ref{Sec:experiandresults} defines implementation features and presents key results. Finally, Section \ref{Sec:discussion} and Section \ref{Sec:conclusion} discuss the future work and the conclusions, respectively.

\section{Background}\label{Sec:background}
This methodology is proposed to study the reputation behaviour in consensus algorithms for blockchain networks, and it is based on intuitionistic fuzzy sets and uniform aggregation operations. Then, this section briefly describes blockchain, intuitionistic fuzzy sets, and uninorm aggregation operations. Before defining the main concepts, the nomenclature used in this work is presented in Table \ref{Tab: nomenclature}. This table provides definitions for key terms to ensure clarity and consistency throughout the text.

\begin{table}[!t]
\centering
\begin{tabular}{cc}
\hline
\multicolumn{2}{c}{\textbf{Nomenclature}} \\\hline
$V$	& Set of validators \\
$V_{\tau}$	& Set of validators $> \tau$ \\
$v_i$	& $i$-th Validator \\
$t$	& Current round \\
$SuccVR(v_i,t)$	& Successful Validation Rate \\ 
$RepD(v_i,t)$	& Reputation Degree \\ 
$w(v_i,t)$ &	Reputation Weight \\ 
$SV(v_i,t)$& Successful Validation\\ 
$A$	& Intuitionistic fuzzy set \\ 
$\mu_A(x)$ &	Membership Function \\ 
$\nu_A(x)$ &	Non-Membership Function \\\ 
$\pi_A(x)$ &	Intuitionistic Index \\ 
$e$	& Neutral element \\ 
$U(x,y)$	& Uninorm aggregated operator \\ 
$T(a,b)$	& Triangular norm \\ 
$S(a,b)$	& Triangular conorm \\\hline
\end{tabular}
\caption{This table shows the nomenclature used in this article.
\label{Tab: nomenclature}}
\end{table}

\subsection{Blockchain}

Blockchain technology was introduced by the pseudonymous Satoshi-Nakamoto in the context of Bitcoin \cite{Nakamoto2008}. A blockchain is a decentralised, immutable, and distributed ledger technology that enables a secure and transparent way to record transactions and verify data through a public, private, or hybrid network of interconnected nodes or computers \cite{Zheng2023}. Blockchains can be seen as distributed databases consisting of a continuously growing list of records, stored in a chain of blocks, which, through cryptographic algorithms, are linked and timestamped to ensure the integrity of the information. The decentralised design ensures that no single entity has control over the entire network, providing resistance to tampering and fault tolerance, preventing the centralisation of power and mitigating the risks of single points of failure.

A crucial concept within the blockchain is the consensus algorithms \cite{Chaudhry2018}. The decentralised nature of this technology requires a consensus mechanism to maintain trust and appropriate network functioning. These algorithms also enable distributed networks of stakeholders to achieve consensus or agreement on the state of the blockchain despite the lack of trust between participants. Moreover, this consensus ensures that only valid and verified transactions are added to the blockchain while preventing malicious actions, such as double-spending \cite{Kumar2023}.

There are several types of consensus algorithms, each suited to different use cases and environments. The two most well-known consensus algorithms are PoW \cite{Nakamoto2008} and PoS \cite{Ethereum2023}.  The former was popularised by  Bitcoin and requires participants, known as miners, to solve complex mathematical puzzles to validate transactions and create new blocks \cite{Nakamoto2008}. The latter was popularised by Ethereum, where the validators are chosen to create new blocks based on the number of ethers they hold and are willing to "stake" as collateral \cite{Ethereum2023}. PoW is known for its security but is criticised for its energy consumption and scalability limitations, as the computational effort required increases with network growth \cite{Wendl2023}. PoS is more energy-efficient than PoW, and both lack diversification, so only a selected group of participants with higher computational power or higher stake is chosen to validate and verify the block transaction \cite{Fahim2023}.  In addition to PoW and PoS, other consensus mechanisms have emerged, such as Delegated Proof of Stake (DPoS) \cite{Larimer2014dpos} and Byzantine Fault Tolerance (BFT) \cite{Castro1999}, each designed to address specific scalability, security, or decentralisation challenges. 

Consensus algorithms are the backbone of blockchain technology, ensuring the trustworthiness and reliability of distributed ledger systems, and their continued development and innovation are key to the evolution of blockchain ecosystems. The development and optimisation of these consensus algorithms remains critical to the widespread adoption and future growth of blockchain applications across various industries, from finance and supply chain management to healthcare and governance \cite{Kumar23future}.

\subsection{Intuitionistic fuzzy sets}
Fuzzy logic offers a powerful framework for dealing with uncertainty and imprecision \cite{Zadeh1965} by allowing for the representation of vague or ambiguous information using linguistic variables and fuzzy sets. 
Intuitionistic fuzzy sets are introduced by Atanassov \cite{Atanassov99} as an extension of Zadeh fuzzy sets \cite{Zadeh1965}, where in addition to the membership function $\mu_A$ defined for fuzzy set $A$, there exists a non-membership function $\nu_A$, such that $\mu_A+\nu_A \in [0,1]$. When $\mu_A+\nu_A=1$, the intuitionistic fuzzy set behaves like a classical fuzzy set, as the degree of membership is the complement of the degree of non-membership. The quantity $1-\mu_A-\nu_A$, known as the intuitionistic index, degree of non-determinacy or degree of uncertainty, is also relevant. A formal definition for the IFS is given in Definition \ref{Def:ifs}.

\begin{definition} \label{Def:ifs}
    \cite{Atanassov99} Let $X$ be a fixed set. An intuitionistic fuzzy set (IFS) $A$ over a set $X$ is defined as
   \begin{equation}
A = \{ (x, \mu_A(x), \nu_A(x)) : x \in X\},
\end{equation}
where the functions 
\begin{equation}
  \mu_A(x)\colon X\rightarrow[0,1]  
\end{equation}
and 
\begin{equation}
    \nu_A(x)\colon X\rightarrow[0,1]
\end{equation}
define the degree of membership and the degree of non-membership of the element $x \in X$, respectively, and for every $x \in X$:
\begin{equation}
    0\leq \mu_A(x)+\nu_A(x) \leq 1.
\end{equation}
The value of 
\begin{equation}
    \pi_A(x)=1-\mu_A(x)-\nu_A(x)
\end{equation}
is called the intuitionistic index, degree of non-determinacy or degree of uncertainty of the element $x\in X$ to the intuitionistic fuzzy set A.
\end{definition}

This methodology uses IFSs to express variable reputation and computes a reputation degree for the nodes, indicating their honesty and providing insights into their reliability. IFSs allow us to separately consider evidence that a node is honest against evidence that is dishonest. For example, consider the case where $\mu_A = 0.8$ and $\nu_A = 0.1$, representing an individual for whom substantial evidence indicates honest behaviour, in contrast to $\mu_A = 0.08$ and $\nu_A = 0.01$, which correspond to an individual about whom little evidence is available, despite most of it suggesting honesty.

IFSs offer a valuable framework for handling uncertain and vague information with a more semantic approach compared to traditional fuzzy sets \cite{Yalcin2025}. IFSs introduce a function that takes into account the degree of non-membership and the intuitionistic index.

\subsection{Uninorm aggregation operations}
Uninorm aggregation operators are a generalisation of the t-norm and t-conorm. Yager \cite{Yager1993, Yager1996} states that the t-norm and t-conorm can characterise the $and$ and $or$ operators used in fuzzy logic, and they were defined as follows.

\begin{definition}\label{Def:tnorm}
    \cite{Yager1993, Yager1996} A triangular norm or t-norm $T$ is a mapping
    \begin{equation}
        T:[0,1]\times[0,1]\rightarrow[0,1]
    \end{equation}
    having the following properties for all $a,b, c, d \in [0,1]$
    \begin{enumerate}
    \item Commutativity: $T(a,b)=T(b,a)$ 
    \item Associativity: $T(T(a,b),c)=T(a,T(b,c))$ 
    \item Monotonicity:  $T(a,b)\leq T(c,d)$ if $a\leq c$ and $b\leq d$
    \item Neutral element: $T(a,1)=a.$
    \end{enumerate}
\end{definition}

\begin{definition}
    \cite{Yager1993, Yager1996} A triangular conorm or t-conorm $S$ is a mapping
    \begin{equation}
        S:[0,1]\times[0,1]\rightarrow[0,1]
    \end{equation}
    having the properties 1-3 in Definition \ref{Def:tnorm} plus
    $$S(a,0)=a.$$
\end{definition}

The uninorm aggregation operations share the initial three properties with both the t-norm and t-conorm but offer greater flexibility concerning the fourth property. t-norms and t-conorms have 1 and 0, respectively, as the neutral element. Uninorms may have any element $e$ in the unit interval as the neutral element. The following definition captures the concept of uninorm.

\begin{definition}\label{Def:uninorm}
    \cite{Yager1996} An uninorm $U$ is a mapping
    \begin{equation}
        U:[0,1]\times [0,1]\rightarrow[0,1]
    \end{equation}
    having the following properties for all $a,b, c, d \in [0,1]$
    \begin{enumerate}
    \item Commutativity: $U(a,b)=U(b,a)$ 
    \item Associativity: $U(U(a,b),c)=U(a,U(b,c))$ 
    \item Monotonicity:  $U(a,b)\leq U(c,d)$ if $a\leq c$ and $b\leq d$
    \item Neutral element: $\exists e \in [0,1]: \forall a \in [0,1]:U(a,e)=a$.
    \end{enumerate}
\end{definition}

Notice that, in Definition \ref{Def:uninorm}, when  $e = 1$ appears, the specific case for the t-norm and when $e = 0$  becomes the t-conorm.

Uninorm aggregation operators are essential in fuzzy logic systems for combining and synthesising uncertain or imprecise information from multiple sources. Moreover, they provide a flexible framework for modelling complex relationships and uncertainties.

\section{Related work}\label{Sec:rw}
Consensus algorithms in the blockchain are crucial mechanisms responsible for establishing trust between the participants in the network and, at the same time, providing security services such as integrity and privacy data. For this reason, some studies have contributed to understanding the behaviour of the participants involved in the consensus algorithms. In particular, reputation-based consensus algorithms have gained attention for their ability to evaluate and integrate the behaviour of network participants into the consensus process, where the trustworthiness of validators is reflected through assigned reputation scores. These algorithms introduce an adaptive approach to validator selection, where reputation can influence which nodes participate in block validation, making the consensus process more secure and efficient. For instance, Bugday et al. \cite{Bugday2019} focused on forming a consensus group using an online learning-based reputation model, which selects nodes with high reputation values. Oliveira \cite{Oliveira2020} presented the Blockchain Reputation-Based Consensus (BRBC) mechanism, which requires a node to have a reputation score above a network trust threshold to insert a new block and uses a set of judges to monitor and update node reputation. Abdo et al. \cite{Abdo2020} introduced a permissionless pure reputation-based consensus algorithm. In contrast, Aluko et al. \cite{Aluko2021} proposed a Proof-of-Reputation (PoR) mechanism that uses a liquid rank algorithm to calculate node reputation. These studies collectively highlight the potential of reputation-based consensus algorithms, nevertheless, each of these works is focused on studying the problem in a specific consensus algorithm primarily focused on threshold-based approaches to reputation management, where reputation scores are often adjusted by external monitoring mechanisms (such as judges or predefined algorithms).  

In this paper, the proposal focuses on a reputation-based consensus algorithm, where a validator's reputation fluctuates according to its behaviour over time. Validators can experience increases or decreases in their reputation based on the quality of their participation in the network, allowing for a more responsive and adaptive system. This approach ensures that validators who demonstrate consistently trustworthy behaviour are rewarded, while those who act maliciously or inefficiently see their reputation decrease, eventually leading to their exclusion from the consensus process if necessary.

To manage this reputation fluctuation effectively, the proposed algorithm integrates fuzzy logic principles, which are well-suited for handling uncertainty and imprecise information in dynamic environments. Prior studies on fuzzy logic in trust management provide a useful foundation. Carbó et al. \cite{Rubiera2003} presented a trust management mechanism that uses fuzzy sets to handle uncertain information about others in electronic commerce.  Quesada et al. \cite{Quesada2015} proposed a methodology to manage the behaviour of experts in large-scale consensus reaching for group decision-making problems, using fuzzy sets and uninorm aggregation operations. However, within the blockchain area, few works integrate these tools into consensus algorithms. Recently, Ramos et al. \cite{Ramos2024} utilised fuzzy sets theory and computing with words in the PoS consensus algorithm to manage the uncertainty that exists in the stake for each validator. As a consequence, the authors designed an equitable consensus mechanism and provided diversification in the selection process to choose the validators. Following the field of fuzzy logic, the author in \cite{Singh2024} proposed a fuzzy-based miner selection algorithm applied to the Internet of Medical Things (IoMT). The algorithm considers parameters such as node status, coin (token) stake, voting, reputation, randomness, and neighbour density.

These studies collectively highlight the growing interest in modelling reputation behaviour in blockchain networks. By allowing reputation to fluctuate based on validator behaviour, the proposed approach enhances network adaptability, fairness, and security, while ensuring and incentivising positive participation. Furthermore, integrating novel tools into blockchain systems, such as intuitionistic fuzzy sets and uninorm aggregation operations, to model reputation, may contribute to the development of more adaptive, efficient, and secure decentralised systems capable of managing uncertain and dynamic environments.


\subsection{Reputation model}

\begin{figure}[!t]
\centering
\includegraphics[width=8.6cm]{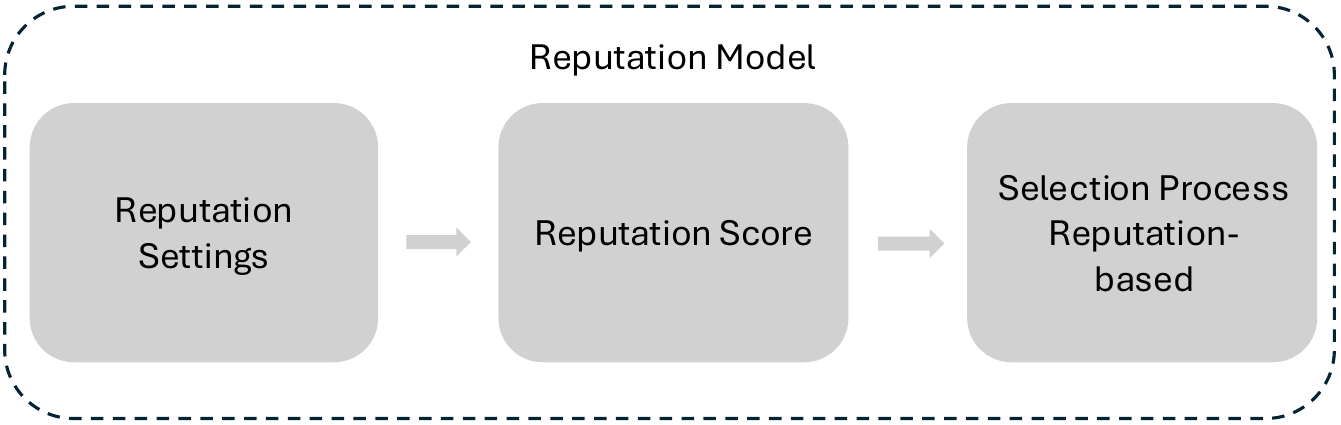}
\caption{The figure illustrates three sequential stages of the structured framework for reputation models.}
\label{Fig:repmodel}
\end{figure}

In the literature, there are various reputation-based consensus algorithms, such as \cite{Abdo2020, Cai20, Chen2024, Oliveira2020, Deng2024, Gan2025, Hussain2025, jia2025, Jiang23, Lei18, Nguyen19, Ramos2024, Sarfaraz23, Wu2024, Yang19, Zhang23,  Zhuang19} that follow a general structured framework of reputation models. Such a framework consists of three sequential stages: reputation settings, reputation scoring, and reputation-based selection (see Figure. \ref{Fig:repmodel}). In the reputation settings stage, the system parameters are configured. For instance, every node in the network is assigned a reputation value $Rep \in [0,1]$, with an initial reputation value $Rep_0$, that is the same for every node. The reputation scoring stage involves the algorithms used to compute each node's reputation score. The algorithm applied to calculate the reputation score depends on the reputation-based consensus algorithm; for example, in \cite{Zhuang19} the reputation algorithm computes the score using the transaction data in the transaction block and the behaviour of nodes. Finally, the reputation-based selection process leverages these scores to prioritise or select nodes for the validation and verification process. 

In the following section, we present our methodology for managing reputation behaviour in reputation-based consensus algorithms for blockchain systems.


\section{Uninorm-based methodology to manage the reputation behaviour in blockchain}\label{Sec:methodology}

\begin{figure}[!t]
\centering
\includegraphics[width=9cm]{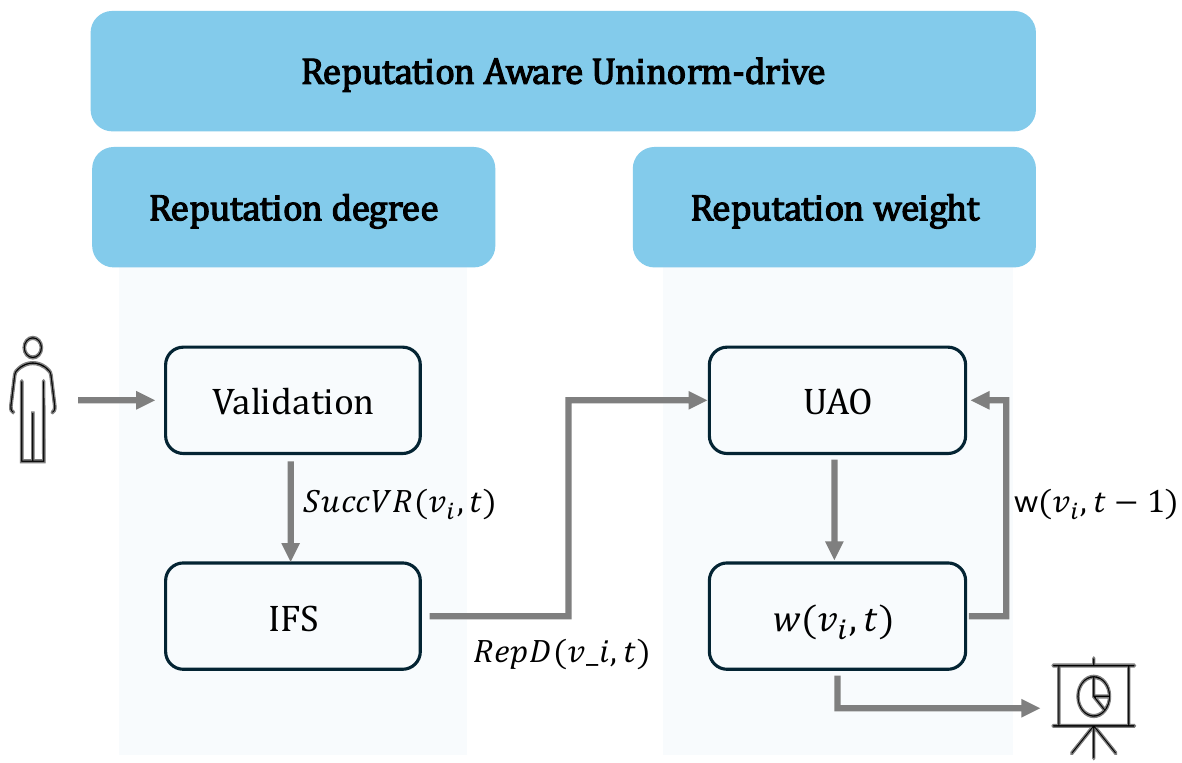}
\caption{Figure illustrates the methodology phases based on IFS and UAO for managing reputation behaviour in the consensus algorithms to the blockchain networks.}
\label{Fig:overviewmetho}
\end{figure}

This section presents our novel proposal, a reputation aware uninorm-driven approach for managing the reputation of each node in the blockchain network. Figure \ref{Fig:reputationmodel} illustrates how our novel proposal, the reputation aware uninorm-driven approach, is integrated into the reputation model for reputation-based consensus algorithms. In the classic reputation model, the reputation selection process typically prioritises the node with the highest reputation in the current round (where a round refers to each time a block is validated). However, our proposed method extends this approach by incorporating not only the reputation from the current round but also the reputation from the previous round. By doing so, our proposal enhances the selection process, enabling more informed and effective decision-making. Therefore, our proposal is structured into two phases, which are depicted in Figure \ref{Fig:overviewmetho}:

\begin{figure}[!t]
\centering
\includegraphics[width=8.5cm]{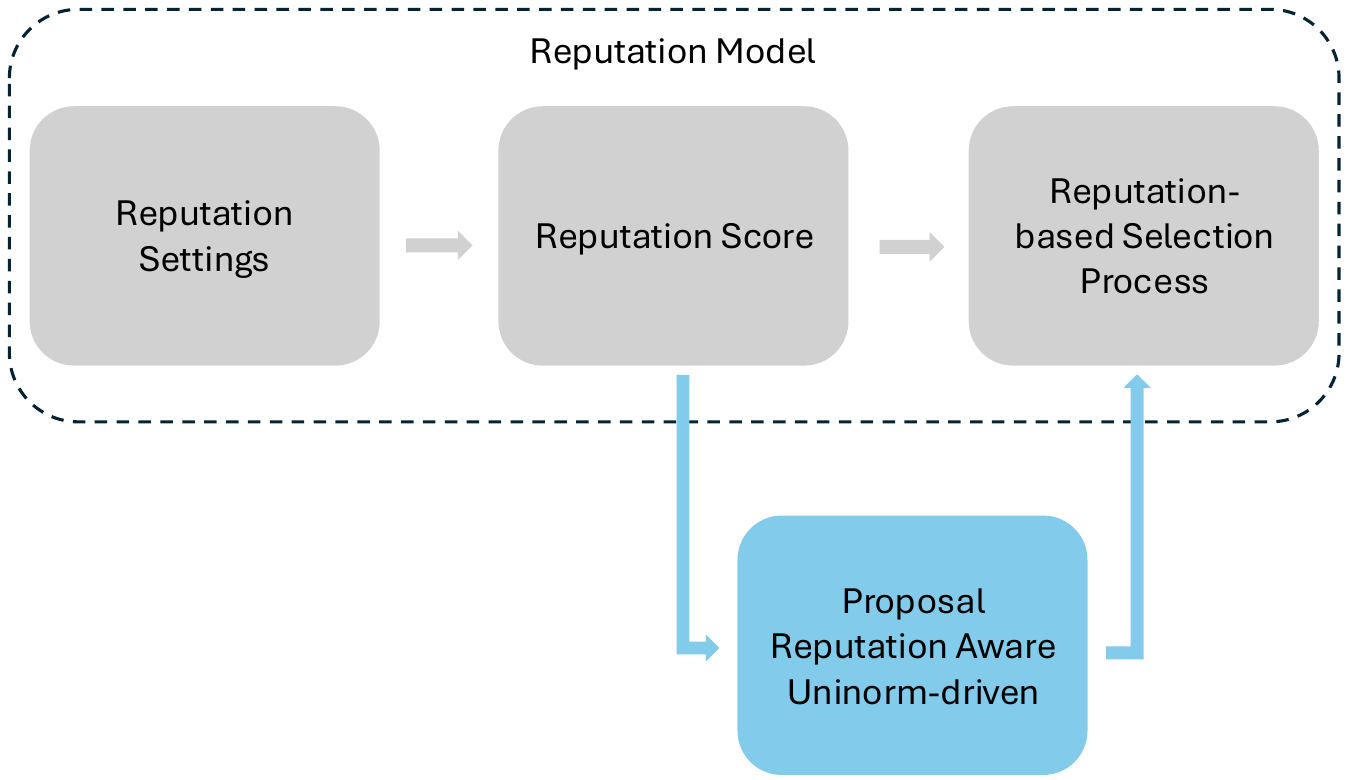}
\caption{Figure illustrates the framework for the integration of our proposal reputation aware uninorm driven into the reputation model in reputation-based consensus algorithms.}
\label{Fig:reputationmodel}
\end{figure}

\begin{itemize}
    \item \textbf{Reputation degree.}  In this phase, the validation process is analysed, and each validator $v_i \in V$ obtains a Successful Validation Rate $SuccVR(v_i,t)$ for the current round $t$. This rate is then used in the Intuitionistic Fuzzy Set (IFS) to compute the Reputation Degree $RepD(v_i,t)$. The  $RepD(v_i,t)$ shows short-term behaviour, which can be used in the algorithm, nevertheless,  calculating only the reputation degree can lead to errors.
    \item \textbf{Reputation weight.} For the next phase, the UAO utilises the $RepD(v_i,t)$ to calculate the Reputation Weight $w(v_i,t)$, taking into account the reputation weight from the previous round $w(v_i,t-1)$. The reputation weight $w(v_i,t)$ shows long-term behaviour, providing greater security. Therefore, the $w(v_i,t)$ is analysed and monitored to study the validator's behaviour during the validation and verification process.
\end{itemize}

Once the reputation weight $w(v_i,t)$ has been computed for all validators, the selection process is performed. First, a predefined threshold $\tau \in [0,1]$ is established to filter eligible validators. Only those validators whose reputation weight satisfies $w(v_i,t) \geq \tau$ are included in the candidate set $V_{\tau} = \{ v_i \mid w(v_i,t) \geq \tau \}$. This threshold mechanism ensures that only validators demonstrating sufficient long-term trustworthy behaviour are considered for participation in the current validation round.

Subsequently, rather than deterministically selecting the validator with the highest reputation weight, a randomised selection algorithm is applied over the candidate set $V_{\tau}$. The random selection may follow a uniform or weight-proportional distribution based on $w(v_i,t)$, depending on the desired balance between fairness and performance. This two-stage mechanism enhances security and decentralisation by ensuring that only validators with acceptable reputation levels participate in the block validation and transaction verification process.

\begin{algorithm}
\caption{Reputation aware uninorm-driven}
\label{alg:selection}
\begin{algorithmic}[1]

\Require $V = \{v_1, v_2, ..., v_n\}$, Current round $t$, Threshold $\tau$
\Ensure Selected validator $v_{selec}$

\State $V_{\tau} \gets \emptyset$ 

\For{each $v_i \in V$}
    \State Calculate $SuccVR(v_i, t)$
    \State Calculate $RepD(v_i, t)$ using IFS
    \State Calculate $w(v_i, t)$ using UAO and $w(v_i, t-1)$
    
    \If{$w(v_i, t) \geq \tau$}
        \State $V_{\tau} \gets V_{\tau} \cup \{v_i\}$
    \EndIf
\EndFor

\If{$V_{\tau} = \emptyset$}
    \State Select $v_{selec}$ such that $w(v_{selec},t) = \max\limits_{v_i \in V} w(v_i,t)$
\Else
    \State Randomly select $v_{selec}$ from $V_{\tau}$
    
\EndIf

\State \Return $v_{selec}$
\end{algorithmic}
\end{algorithm}

\begin{proposition}[Liveness and Safety]
If at least one validator satisfies $w(v_i,t)\ge\tau$, the selection step
guarantees liveness by selecting a validator in finite time. Furthermore,
since $w(v_i,t)$ encodes long-term behaviour, the probability of selecting a
faulty validator decreases exponentially with consecutive failures.
\end{proposition}

It is important to note that the proposed selection step is modular and can be incorporated into various reputation-based consensus protocols. In particular, it can replace stake-based probability mechanisms similar to PoS schemes, while the reputation update procedure remains as defined in Section \ref{Sub:reputationdegre}. Reputation degree and Section \ref{Sub:reputationweight}. Reputation weight.

\subsection{Reputation degree} \label{Sub:reputationdegre}
The first step in computing the reputation degree is to calculate the successful validation rate. During the validation process, each participant has a successful validation rate, as presented in Definition \ref{Def:succesvr}. This rate is the number of successful validations divided by the total number of validation attempts. 

\begin{definition}\label{Def:succesvr}
    Let $\#SV$ be the number of successful validations. Then the successful validation rate, $SuccVR(v_i,t)$ by each validator $v_i$ in the current round $t$ is defined by the function:
\begin{equation}\label{Eq:Succesvr}
    SuccVR(v_i,t)=\frac{\#SV(v_i,t)}{t}
\end{equation}
\end{definition}

\begin{remark}
    The number of validation attempts corresponds to the number of rounds. In each round, the validator has only one attempt. 
\end{remark}

The successful validation rate will be used to compute the reputation degree through an intuitionistic fuzzy set $A$ that expresses the linguistic variable ``reputation". 

\begin{figure}[!t]
\centering
\includegraphics[width=8.5cm]{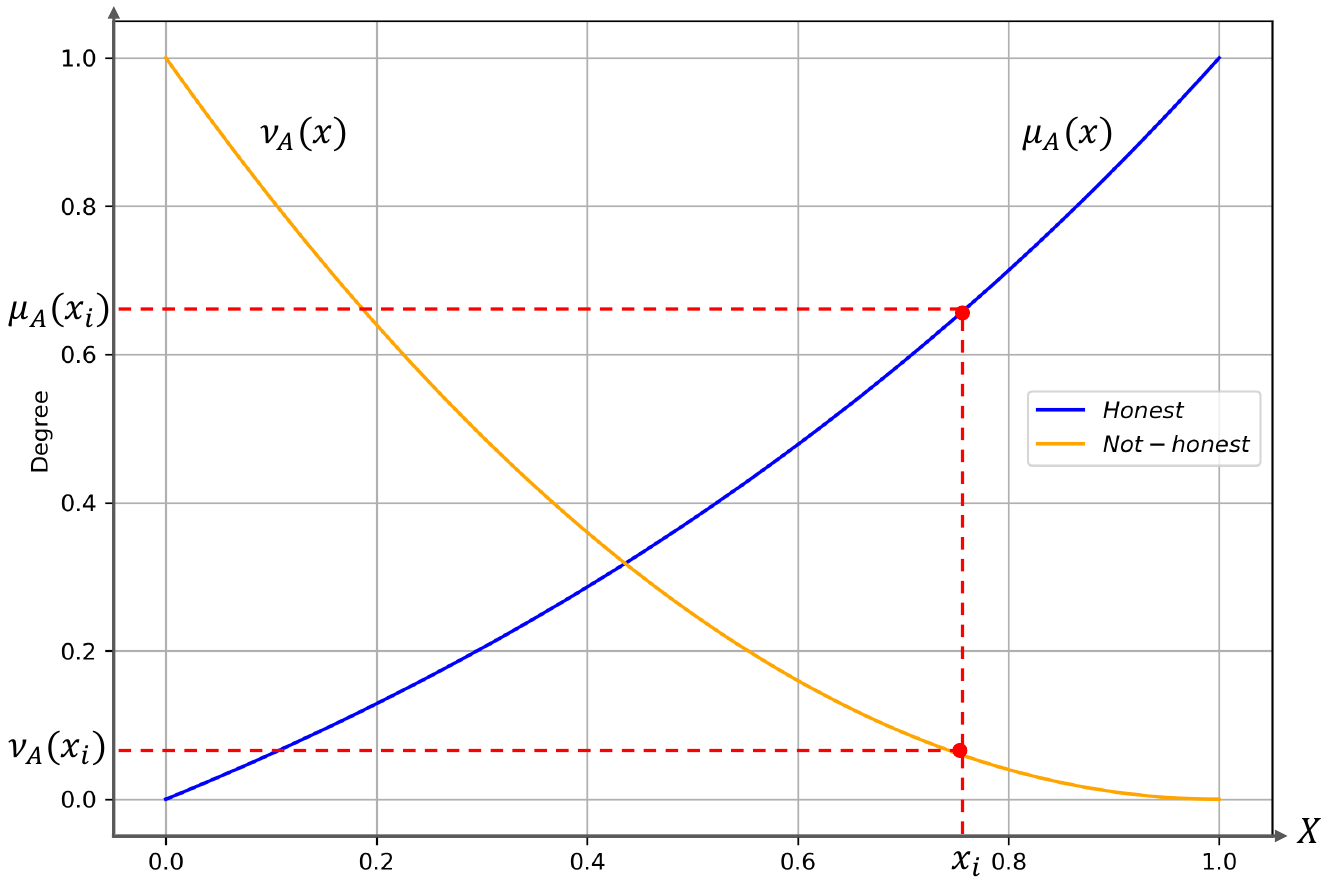}
\caption{Figure shows the intuitionistic fuzzy set $A$ with the membership function $\mu_A(x)$ and the non-membership function $\nu_A(x)$.}
\label{Fig:ifsA}
\end{figure}

According to Definition \ref{Fig:ifsA}, an intuitionistic fuzzy set $A$ assigns to each element $x$ of the universe $X$ a membership degree $\mu_A(x) \in [0,1]$ and a non-membership degree $\nu_A(x) \in [0,1]$ such that:
$0 \leq \mu_A(x) + \nu_A(x) \leq 1$. Moreover, for all $x \in X$ recall that $\pi_A(x)= 1-\mu_A(x)- \nu_A(x)$. This paper focuses on studying the reputation of validators in the consensus algorithm in the blockchain. Therefore, as the universe of discourse represents the successful validation rate  $SuccVR(v_i, t)$ defined on the real interval $X = [0,1]$, following the Definition \ref{Def:succesvr}. 

The membership function $\mu_A(x)$ for $x= SuccVR(v_i, t) \in A$ is defined as:

\begin{eqnarray}\label{Eq:muAx}
  \mu_A(x)= \begin{cases}
 0 & \text{ if } x=0  \\ 
 \mu(x) & \text{ if } 0 < x < 1 \\ 
 1 &  \text{ if } x=1
\end{cases}   
\end{eqnarray}
where $\mu_A(x):X \rightarrow[0,1]$ defines the rate at which a node’s reputation increases based on its behaviour over time (see Figure \ref{Fig:ifsA}). The design of $\mu(x)$ is guided by the principle of determining how quickly a validator can regain its reputation after exhibiting undesirable behaviour, such as failing to validate a block correctly. For example, a minor failure might allow for quicker recovery, while repeated or severe infractions could result in a slower pace.  Different options for defining function $\mu(x)$ include:
\begin{enumerate}
    \item Constant increase in reputation: In this case, a linear function is suitable.
    \item Slow increase in reputation: Logarithmic and square root functions may be useful.
    \item Rapid increase in reputation: Exponential and polynomial functions can be applied.
    \item Initial rapid increase in reputation followed by gradual slowing: For instance, logistic and hyperbolic functions.
    \item Slow initial increase in reputation followed by a rapid growth: In this scenario, a sigmoid function could be an option.
\end{enumerate} 

\begin{figure}[!t]
\centering
\includegraphics[width=8.5 cm]{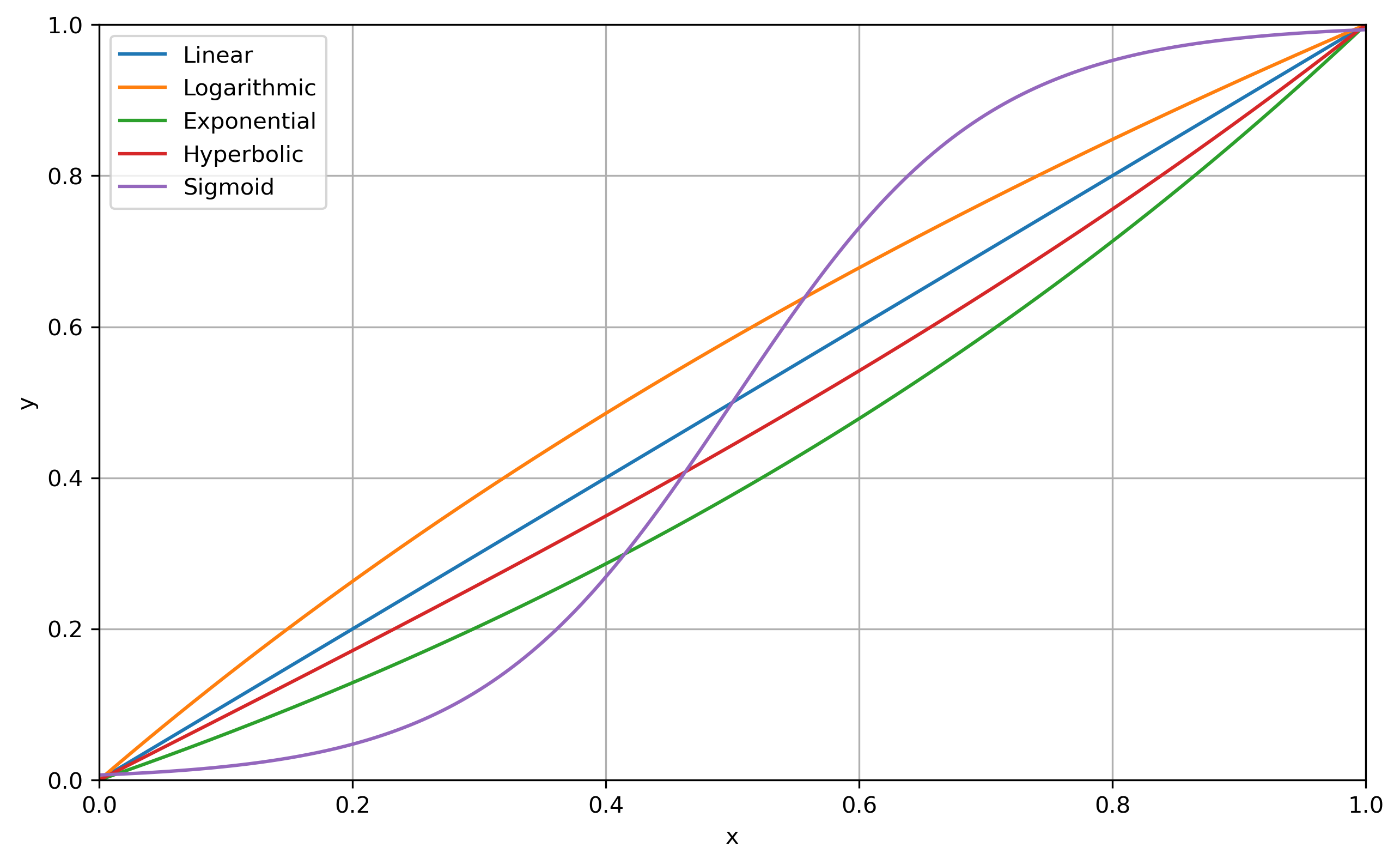}
\caption{Figure illustrates a set of functions that could be used to define the function $\mu(x)$ }
\label{Fig:mus}
\end{figure}

 The list provided above illustrates several examples of how the function $\mu(x)$ can be defined to reflect the enhancement of a validator's reputation after encountering failures as shown in Figure \ref{Fig:mus}. However, it is important to note that these are merely illustrative examples, and other functions can be utilised to define $\mu(x)$. The choice of function depends on specific factors, such as the desired rate of reputation increase.

\begin{figure}[!t]
\centering
\includegraphics[width=8.5 cm]{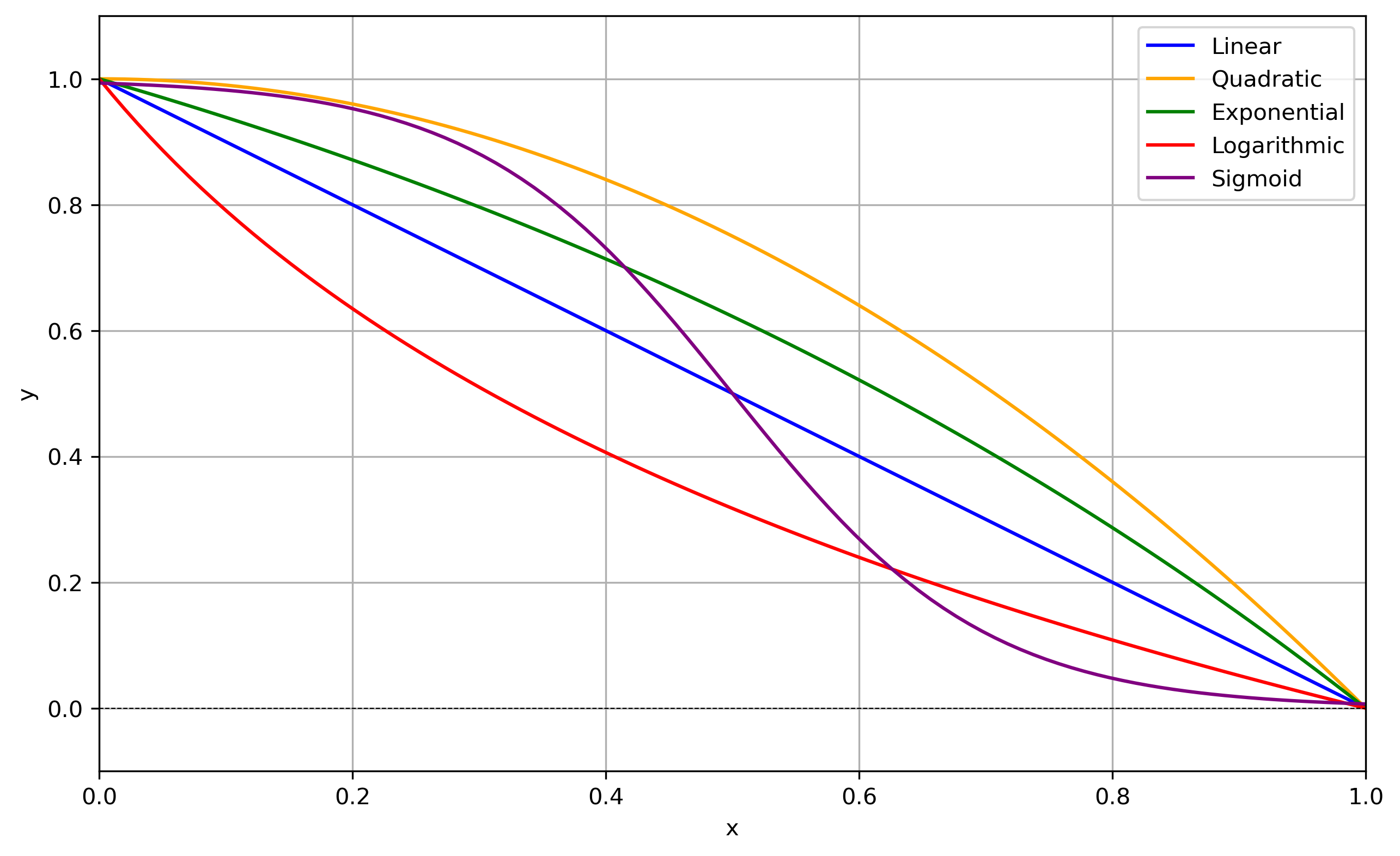}
\caption{Figure illustrates a set of functions that could be used to define the function $\nu(x)$ }
\label{Fig:nus}
\end{figure}

With regard to the non-membership function $\nu_A(x)$ for $x \in A$ is defined as follows:

\begin{eqnarray}\label{Eq:nuAx}
  \nu_A(x)= \begin{cases}
 1 & \text{ if } x=0  \\ 
 \nu(x) & \text{ if } 0 < x < 1 \\ 
 0 &  \text{ if } x=1
\end{cases}   
\end{eqnarray}

where $\nu_A(x):X\rightarrow[0,1]$ is a function defined considering a similar philosophy that was used to define $\mu(x)$ in Eq. (\ref{Eq:muAx}); however,  in this case, the desired rate of decrease is analysed. For instance, if a consistent penalisation is required, a linear function may be employed. Alternatively, opting for a negative logarithmic function results in a rapid initial decrease followed by a slower decline, which may be suitable for scenarios requiring swift but controlled reputation adjustment (see Figure \ref{Fig:nus}).

\begin{figure}[!t]
\centering
\includegraphics[width=8.5 cm]{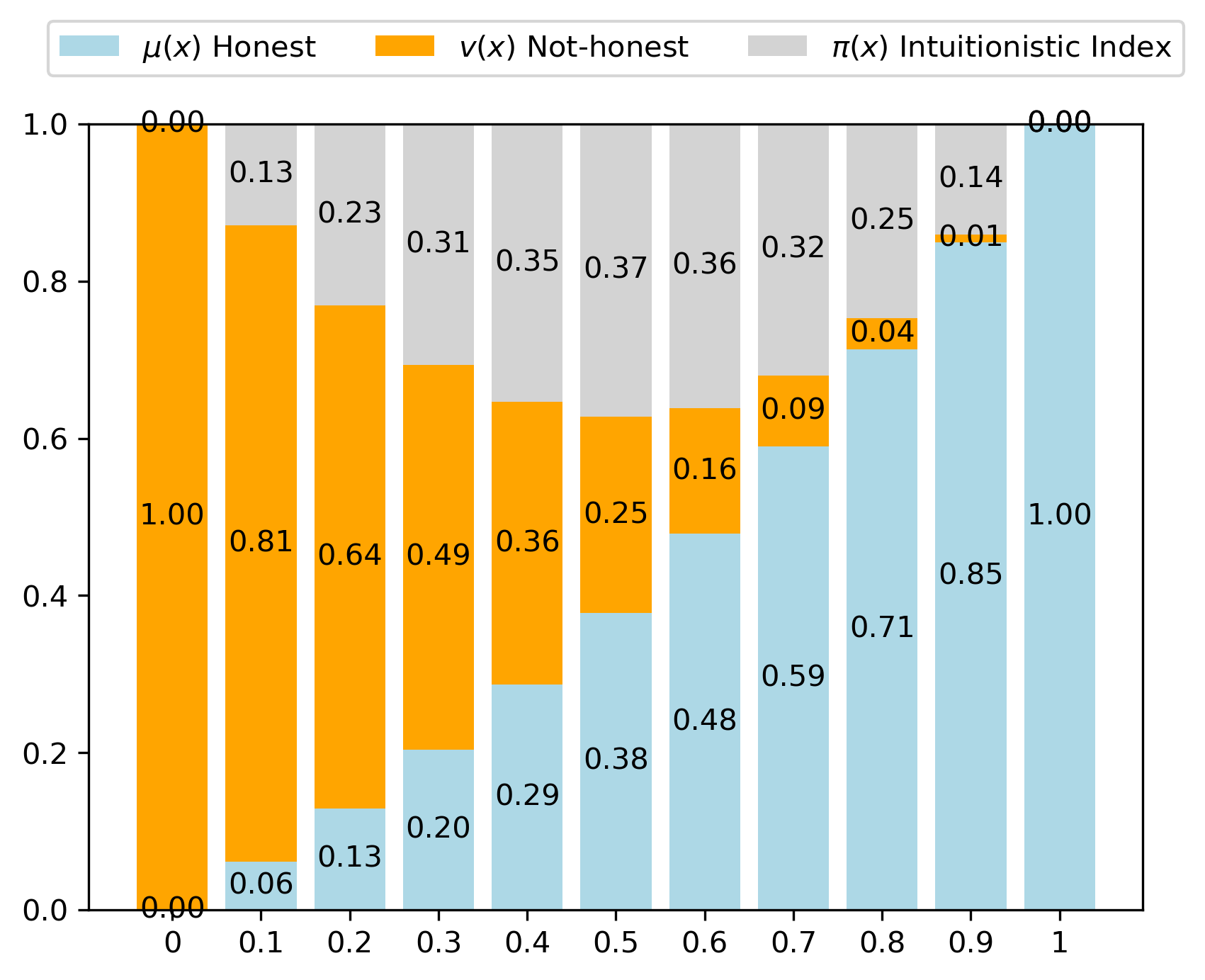}
\caption{Figure displays $\mu_A(x)$, $\nu_A(x)$, and $\pi_A(x)$. }
\label{Fig:munu}
\end{figure}

Reputation is intrinsically a concept filled with uncertainty. It does not constitute an absolute measure but rather emerges from varying perceptions of an entity's behaviour. When new validators join the pool, the available information about their behaviour is incomplete, resulting in inherent indeterminacy, or epistemic uncertainty, regarding the extent to which they are reputable (true) or non-reputable (false). This initial state of indeterminacy evolves dynamically as the blockchain network expands and validators process more transactions. Consequently, an intuitionistic framework for managing uncertainty, one that not only captures degrees of truth (membership) and falsity (non-membership) but also incorporates the dimension of indeterminacy, offers a particularly apt methodology for addressing this context. We employ an intuitionistic index $\pi_A(x)$ to indicate the degree of indeterminacy of the element $x$ to the intuitionistic fuzzy set $A$. Figure \ref{Fig:munu} shows how $\pi_A(x)$ varies with respect to $\mu_A(x)$ and $\nu_A(x)$.

Once the reputation is modelled through the IFS, as shown in Figure \ref{Fig:ifsA}, the next step is to compute the reputation degree for each validator. A natural first candidate for a reputation degree would be the ratio
\[
r(x) = \frac{\mu_A(x)}{\mu_A(x)+\nu_A(x)},
\]
which represents the balance between positive and negative evidence and can be interpreted as a basic reputation rating in the interval $[0,1]$. However, this ratio alone presents a limitation, different combinations of $\mu_A(x)$ and $\nu_A(x)$ may produce the same value of $r$, even though the amount of available information differs. Therefore, this expression does not capture the degree of certainty associated with the evaluation. To address this issue, we incorporate the intuitionistic index $\pi_A(x)$, which quantifies the level of indeterminacy reflecting how much uncertainty remains in the assessment. When $\pi_A(x)$ is small, the evaluation is supported by more information, whereas larger values indicate greater indeterminacy.

Instead of directly adopting $r$ as the reputation degree, we scale it between a lower bound $r_0$ and $r$, where $0<r_0<r$. The parameter $r_0$ prevents unreliable extreme values when the certainty level is low. The scaling depends on a function
\[
s:[0,1]\rightarrow[0,1],
\]
such that $s(0)=0$ and $s(1)=1$. Consequently, when the certainty level is minimal, the reputation degree approaches $r_0$, and when certainty is maximal, it converges to $r$. In this work, $s(\pi_A(x))=\pi_A(x)^\alpha$ with $\alpha>0$, which allows us to regulate the sensitivity of the transition. If $0<\alpha<1$, the model moves rapidly toward $r$ even under moderate certainty; if $\alpha>1$, it remains close to $r_0$ until high certainty is achieved.

These observations are summarised in Definition \ref{Def:repd}, which formally introduces the reputation degree $RepD(v_i,t)$ within the proposed intuitionistic fuzzy framework.

\begin{definition}\label{Def:repd}
    Let A be an intuitionistic fuzzy set, $\mu_A(x)$ the membership degree, $\nu_A(x)$ the non-membership degree, and $\pi_A(x)$ the intuitionistic index, then the reputation degree, $RepD(v_i,t)$, by each validator $v_i$ in the current round $t$ is defined as: 
    \begin{eqnarray}\label{Eq:repd}
  \begin{matrix}
RepD(v_i,t)=r_0\left (\frac{\mu_A(x)}{r_0(\mu_A(x)+\nu_A(x))}  \right )^{(\pi_A(x))^\alpha}, & 
\end{matrix}
\end{eqnarray}
where $0 < r_0 < \frac{\mu_A(x)}{\mu_A(x)+\nu_A(x)}$ and $\alpha>0$.
\end{definition}

Equation \ref{Eq:repd} given in Definition \ref{Def:repd} introduces the reputation degree as a certainty-adjusted of the basic reputation rating derived from the IFS components. To validate that this formulation behaves consistently with its intended interpretation, it is necessary to examine its fundamental mathematical properties. In particular, the reputation degree should remain bounded, respond positively to increasing membership, negatively to increasing non-membership, and exhibit regular behaviour. These properties are formalised in the following proposition.
\begin{proposition}
Let $\mu_A(x)$, $\nu_A(x)$ and $\pi_A(x)$ be the membership, non-membership and intuitionistic index of the IFS, and let $\mathrm{RepD}(v_i,t)$ be given by Equation (12). Then:
\begin{enumerate}
\item $\mathrm{RepD}(v_i,t)\in(0,1)$.
\item $\mathrm{RepD}(v_i,t)$ is monotonically increasing in $\mu_A(x)$.
\item $\mathrm{RepD}(v_i,t)$ is monotonically decreasing in $\nu_A(x)$.
\item $\mathrm{RepD}(v_i,t)$ is continuous on $[0,1]$.
\end{enumerate}
\end{proposition}

For a better understanding of Definition 6,  a numerical illustration is presented in the next paragraph.

\begin{description}
   \item Suppose a validator has successful validation $SuccVR = 0.75$ and the membership and non-membership functions are defined by the following linear IFS functions:
$$\mu_A(x) = x-0.05,\: \nu_A(x) = 0.90-x \: \Rightarrow \: \pi_A(x) = 1- (\mu_A(x)+ \nu_A(x))
$$
Let $\mu_A(0.75) = 0.70$, $\nu_A(0.75) = 0.15$, then $\pi_A(0.75) = 0.15$. Moreover, for this example $r_0 = 0.30$.

Then $$ \frac{\mu_A(x)}{r_0(\mu_A(x)+\nu_A(x))} = \frac{0.70}{0.30(0.70+0.15)} = \frac{0.70}{0.26} \approx 2.6923 $$
Now compute the reputation degree for different $\alpha$:
\begin{itemize}
    \item If $\alpha = 2$: \[ RepD = 0.30 \times (2.6923)^{0.15^2} = 0.30 \times 1.0225 \approx 0.3068 \] 
    \item If $\alpha = 1$: \[ RepD = 0.30 \times (2.6923)^{0.15^1} = 0.30 \times 1.1602 \approx 0.3480 \]
    \item If $\alpha = 0.5$: \[ RepD = 0.30 \times (2.6923)^{0.15^{0.5}} = 0.30 \times 1.4675 \approx 0.4402 \]
\end{itemize}

This example shows that when $\alpha$ is higher the reputation degree decrease, while lower  $\alpha$ the reputation degree increases. This observation is formalised in Remark \ref{Remark:2}.
\end{description}

\begin{remark}\label{Remark:2}
The function given in Definition \ref{Def:repd} can be expressed as 
\[
RepD = r_0 \left(\frac{r}{r_0}\right)^{s(\pi)},
\]
where $r(x)=\frac{\mu_A(x)}{\mu_A(x)+\nu_A(x)}$, $0<r_0<r$. Additionally, the parameter $r_0$ is chosen as a function of $r$ in order to encode how strict the system behaves under uncertainty. For instance, one may define $r_0=\beta r$ with $0<\beta<1$, so that the reputation degree is proportionally reduced when indeterminacy is high. This choice directly relates the baseline value $r_0$ to the underlying membership/non-membership pair $(\mu_A(x),\nu_A(x))$, since $r$ itself is determined by their balance. Consequently, $r_0$ controls how severely the system penalises uncertain evaluations while preserving consistency with the original fuzzy evidence.
\end{remark}

A particular case from Eq. \ref{Eq:repd}, when $\mu_A(x)= c \cdot \nu_A(x)$ with $c=1$: 

  \begin{align*}
RepD(v_i,t)  &= r_0\left (\frac{\mu_A(x)}{r_0(\mu_A(x)+\mu_A(x))}  \right )^{\pi_A(x)^\alpha}  \\ 
 &= r_0\left (\frac{\mu_A(x)}{r_0(2\mu_A(x)}  \right )^{\pi_A(x)^\alpha}\\
 &= r_0\left (\frac{1}{2r_0}  \right )^{\pi_A(x)^\alpha}
\end{align*}
That is semantically appealing, as it suggests ``There is some evidence that you are honest, but also some evidence that you are dishonest, which makes me less sure of your honesty”. 

In the general case, if  $\mu_A(x)= c \cdot \nu_A(x)$ with $0\leq c \leq 1$, then 

\begin{align}
RepD(v_i,t)  &= r_0\left (\frac{\mu_A(x)}{r_0(\mu_A(x)+\frac{\mu_A(x)}{c})}  \right )^{\pi_A(x)^\alpha} \\ 
 &= r_0\left (\frac{\mu_A(x)\cdot c}{r_0(\mu_A(x) \cdot c+\mu_A(x)}  \right )^{\pi_A(x)^\alpha}\\
 &=r_0\left (\frac{ c}{r_0(c+1)}  \right )^{\pi_A(x)^\alpha}
\end{align}

Observe that if $\pi_A(x) \rightarrow 0$, then $RepD(v_i, t)$ rapidly converges to $r_0$, even under high uncertainty. Conversely, when $\pi_A(x) \rightarrow 1$, $RepD(v_i, t)$ remains close to $\frac{c}{c+1}$ until a high level of certainty is achieved. The parameter $\alpha$ allows the behaviour of the function to be tuned: values of $\alpha$ close to zero imply that $RepD(v_i, t)$ depends primarily on the relative magnitudes of $\mu_A(x)$ and $\nu_A(x)$, leading to an increase in the reputation degree, whereas larger values of $\alpha$ make it more dependent on the absolute magnitude of $\mu_A(x)$, resulting in a decrease in the reputation degree.

\subsection{Reputation weight}\label{Sub:reputationweight}

\begin{figure}[!t]
\centering
\includegraphics[width=7.7cm]{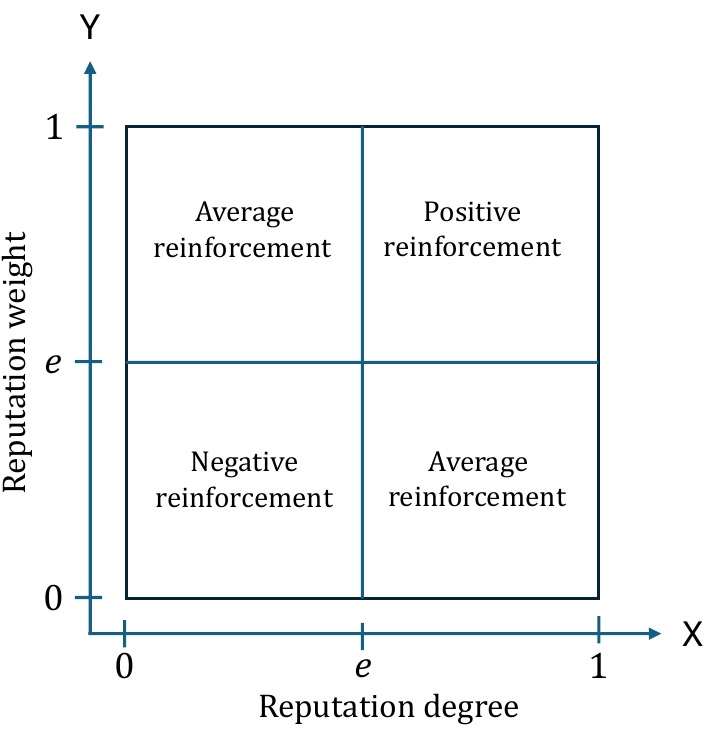}
\caption{Structure of uninorm: reputation weight for different types of reputation degree.}
\label{Fig:structure}
\end{figure}

In this section, we employ the reputation degree computed in the preceding section to determine the corresponding reputation weight. This methodology proposes using an uninorm aggregation operation to model reputation behaviour; through these operators, the reputation degree is aggregated to obtain a reputation weight. The reputation degree and reputation weight are used to monitor and demonstrate the importance of reinforcement. For instance, if the validator $t_i$ exhibits both a high reputation degree and a high reputation weight, it receives positive reinforcement. Conversely, if the validator has both a low reputation degree and a low reputation weight, it receives negative reinforcement. Otherwise, the validator may receive average reinforcement, as shown in Figure \ref{Fig:structure}.

The following definition is provided for computing the reputation weight:

\begin{definition}\label{Def:repweight}
    Let $U$ be the uninorm aggregation operator with the associated neutral element $e$. Then, the reputation weight $w(v_i,t)$, by each validator $v_i$ at the round $t$ is computed as follows:
    \begin{eqnarray}
    w(v_i,t)= \begin{cases}
 e& \text{ if }t=1 \\ 
 U(RepD(v_i,t),w(v_i,t-1))&  \text{ if } t>1
\end{cases}
\end{eqnarray}
\end{definition}
 
Note that $U$ in Definition \ref{Def:repweight} can be defined considering the structure of an uninorm shown in Figure \ref{Fig:structure}. For instance, the operator $U$ can be defined as a piecewise function where a t-norm operator is used if both the reputation degree and the reputation weight are less than the neutral element $e$, a t-conorm is utilised if both the reputation degree and the reputation weight are greater than the neutral element $e$ and less than 1. In other cases, the arithmetic mean between the reputation degree and the reputation weight can be applied. As a particular example, the following Fodor's uninorm \cite{Fodor1997} is shown:

\begin{eqnarray}
U(x,y)= \begin{cases}
 2xy & \text{ if }0 \leq x,y \leq e \\ 
 2(x+ y - 2xy-e) & \text{ if } e \leq x,y \leq 1 \\ 
 \frac{x+y}{2}&  \text{ if } \min(x,y)< e <   \\
 & \hspace{16mm}  \max(x,y)
\end{cases}
\end{eqnarray}

\begin{figure}[!t]
\centering
\includegraphics[width=7.7cm]{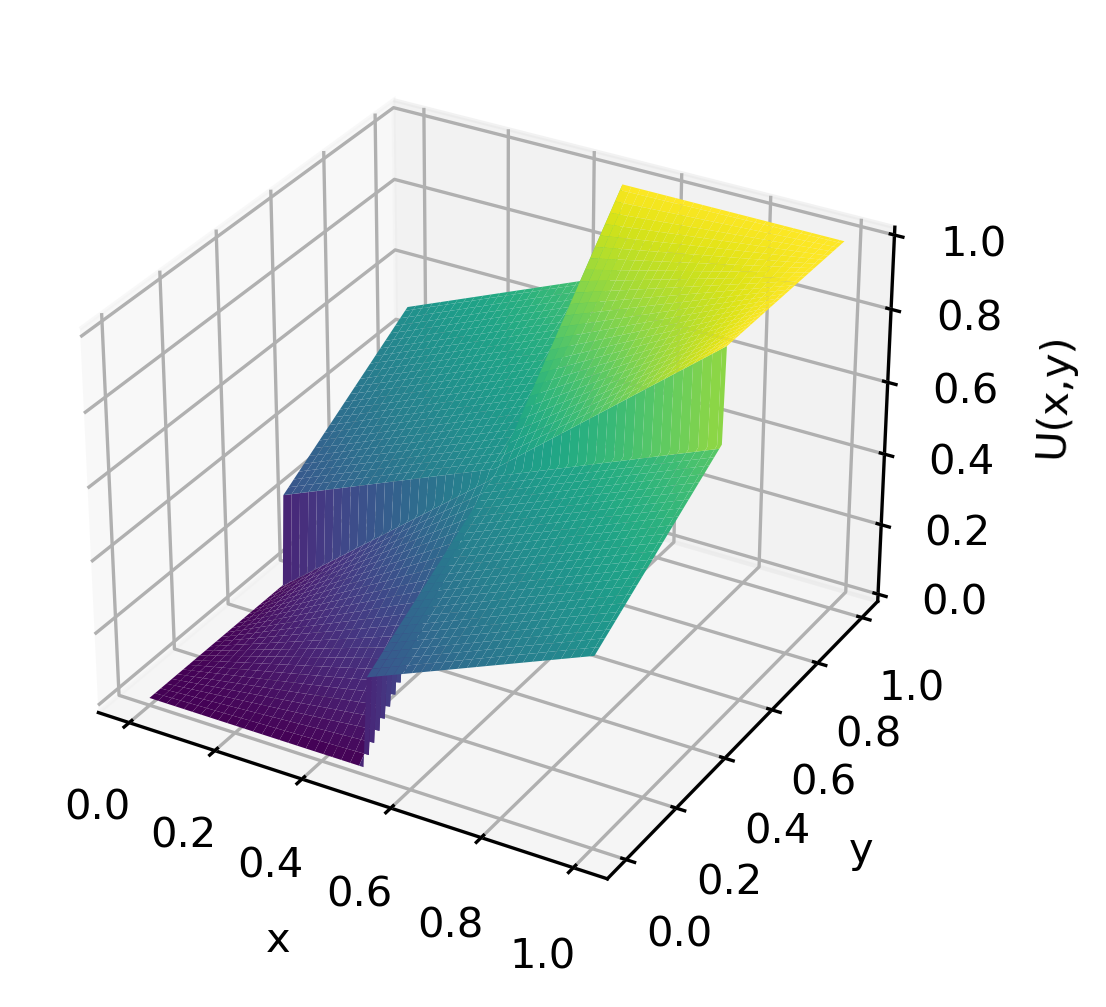}
\caption{Fodor's uninorm with $e=0.5$}
\label{Fig:fodor}
\end{figure}

Figure \ref{Fig:fodor} describes the graphic behaviour of Fodor's uninorm with a neutral element set to $e=0.5$. In all experiments, we adopt the following uninorm with neutral element $e$:
\[
U(a,b) =
\begin{cases}
T(a,b) / e, & \text{if } a,b \le e,\\[4pt]
1 - \dfrac{S(1-a,1-b)}{1-e}, & \text{if } a,b > e,
\end{cases}
\]
where $T$ is the minimum t-norm and $S$ is the maximum t-conorm.

By following this methodology, blockchain networks can leverage uninorm aggregation operators to incorporate information about validator behaviour. The proposed approach can be integrated into consensus algorithms in which participant selection relies on fluctuating reputation as a key parameter. To demonstrate the use of intuitionistic fuzzy sets and uninorm aggregation operators, the next section presents experimental results and discusses the performance of the proposed methodology.


\section{Experiments and results} \label{Sec:experiandresults}
This section performs and discusses a set of experiments designed to assess the performance of the proposed methodology. The experiments were developed using the following software and computer specifications. It includes a CPU and an Intel® Core™ i7-7500U processor, featuring a clock speed of 2.70GHz and four cores. The operating system used is Ubuntu 22.04.3. For compiling, A C++ compiler, GCC version 7.4.0 is utilised. The programming language employed is Python, specifically version 3.10.12.  

\subsection{Experimental results: General evaluation}

The objective of this experiment is to depict the performance of the methodology through the illustrative example of defining and applying specific functions for each phase. According to this methodology, the first phase consists of computing the reputation degree. To do so, simulations have been developed, and
the successful validation rate is computed in each round using the function defined in Eq. (\ref{Eq:Succesvr}). Table \ref{Tab:main} shows the successful validation rate for the validators $v_1$ and $v_2$ in the rows $SucVR(v_1,t)$ and $SucVR(v_2,t)$ over seven rounds. Notice that the validator $v_1$ makes a mistake in rounds 2 and 5, while the validator $v_2$ was wrong in rounds 2 and 3.

\begin{table}[ht]
\centering
\begin{tabular}{ccccccccl} 
\hline
Round(t) & 1 & 2 & 3 & 4 & 5& 6&7\\\hline
$SuccVR(v_1, t)$ & 1.00 & \textbf{0.50} & 0.67 & 0.75 & \textbf{0.60} & 0.67 & 0.71\\
$RepD(v_1, t)$ & 0.62 & 0.40 & 0.55 & 0.59 & 0.50 & 0.55 & 0.58\\
$w(v_1, t)$ & 0.50 & 0.40 & 0.48 & 0.53 & 0.52 & 0.56 & 0.63\\\hline
$SuccVR(v_2, t)$ & 1.00 & \textbf{0.50} & \textbf{0.33} & 0.50 & 0.60 & 0.67 & 0.71\\
$RepD(v_2, t)$ & 0.62 & 0.40 & 0.22 & 0.40 & 0.50 & 0.55 & 0.58\\
$w(v_2, t)$ & 0.50 & 0.40 & 0.18 & 0.14 & 0.14 & 0.35 & 0.46\\
\hline
\end{tabular}
\caption{Table shows the successful validation rate $SucVr()$, the reputation degree $RepD()$, and the reputation weight $w()$ for the validator $v_1$ and $v_2$ at seven rounds.} \label{Tab:main}
\end{table}

The next step is to compute the reputation degree for $v_1$ and $v_2$, considering the successful validation rate $SuccVR$, respectively. For this example, the membership function $\mu_A(x)$ presented in Eq. \ref{Eq:muAx} is defined as:

\begin{eqnarray}
  \mu_A(x)= \begin{cases}
 0 & \text{ if } x<0  \\ 
 \frac{\exp(x)-1}{\exp(1)-1} & \text{ if } 0 \leq x \leq 1 \\ 
 1 &  \text{ if } x>1
\end{cases}   
\end{eqnarray}
where the function $\mu(x)=\frac{\exp(x)-1}{\exp(1)-1}$.

The non-membership function $\nu_A(x)$ in Eq. \ref{Eq:nuAx} is defined by:

\begin{eqnarray}
  \nu_A(x)= \begin{cases}
 1 & \text{ if } x<0  \\ 
 (x-1)^2 & \text{ if } 0 \leq x \leq 1 \\ 
 0 &  \text{ if } x>1
\end{cases}   
\end{eqnarray}
where the function $\nu(x)=(x-1)^2$.

Then, the functions in Eq 18 and Eq. 19 are used together with the function in Eq. 12 to compute the reputation degree for the validator $v_1$ and $v_2$ where the parameter is set to $\alpha=2$ and $r_0=5/8 \cdot \frac{\mu_A(x)}{\mu_A(x)+\nu_A(x)}$ for this example. Table \ref{Tab:main} depicts the outcomes at the row named $RepD(v_1,t)$ and $RepD(v_2,t)$ for $v_1$ and $v_2$, respectively. 

At this moment, the first phase is finished. To start with, the second phase is necessary to define the uninorm aggregation operation $U$. For this example, we will use Fodor's uninorm defined in Eq. 17.  After applying the function presented in Eq. 16, the reputation weight $w(v_i,t)$ is calculated for the validators $v_1$ and $v_2$. The results are presented in the rows $w(v_1,t)$ and $w(v_2,t)$ in Table \ref{Tab:main}.  

\begin{figure}[!t]
\centering
\includegraphics[width=9cm]{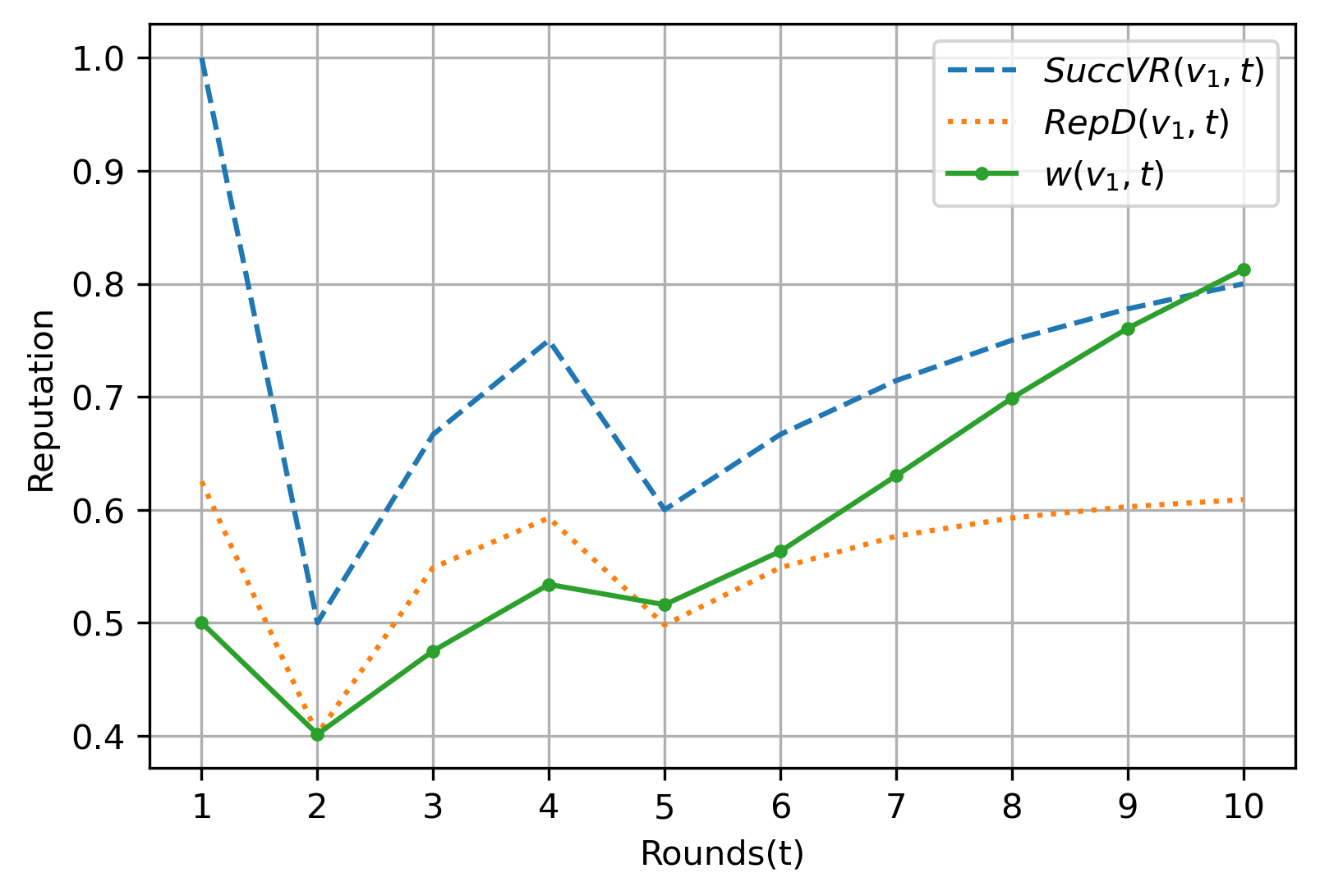}
\caption{Figure plots the results obtained in Table \ref{Tab:main} for the validator $v_1$}
\label{Fig:v1}
\end{figure}

The outcomes of Table \ref{Tab:main} for the validator $v_1$ are displayed in Figure \ref{Fig:v1}. This figure illustrates the reputation behaviour for the validator $v_1$ in ten rounds. The blue graph (dashed line) represents the successful validation rate $SuccVR(v_1,t)$, the orange graph (dotted line) indicates the reputation degree $RepD(v_1,t)$, and the green graph (continuous line) shows the reputation weight $w(v_1,t)$. From Figure \ref{Fig:v1}, it is observed that the validator $v_1$ has a mistake in rounds 2 and 5; thus, the $SuccVR(v_1,t)$, $RepD(v_1,t)$, and $w(v_1,t)$ decrease. For round 3, the validator starts to perform the process correctly, then $SuccVR(v_1,t)$, $RepD(v_1,t)$, and $w(v_1,t)$ increase until round 4, where they decrease again. After round 6, $SuccVR(v_1,t)$, $RepD(v_1,t)$, and $w(v_1,t)$ start to increase again. 

\begin{figure}[!t]
\centering
\includegraphics[width=9cm]{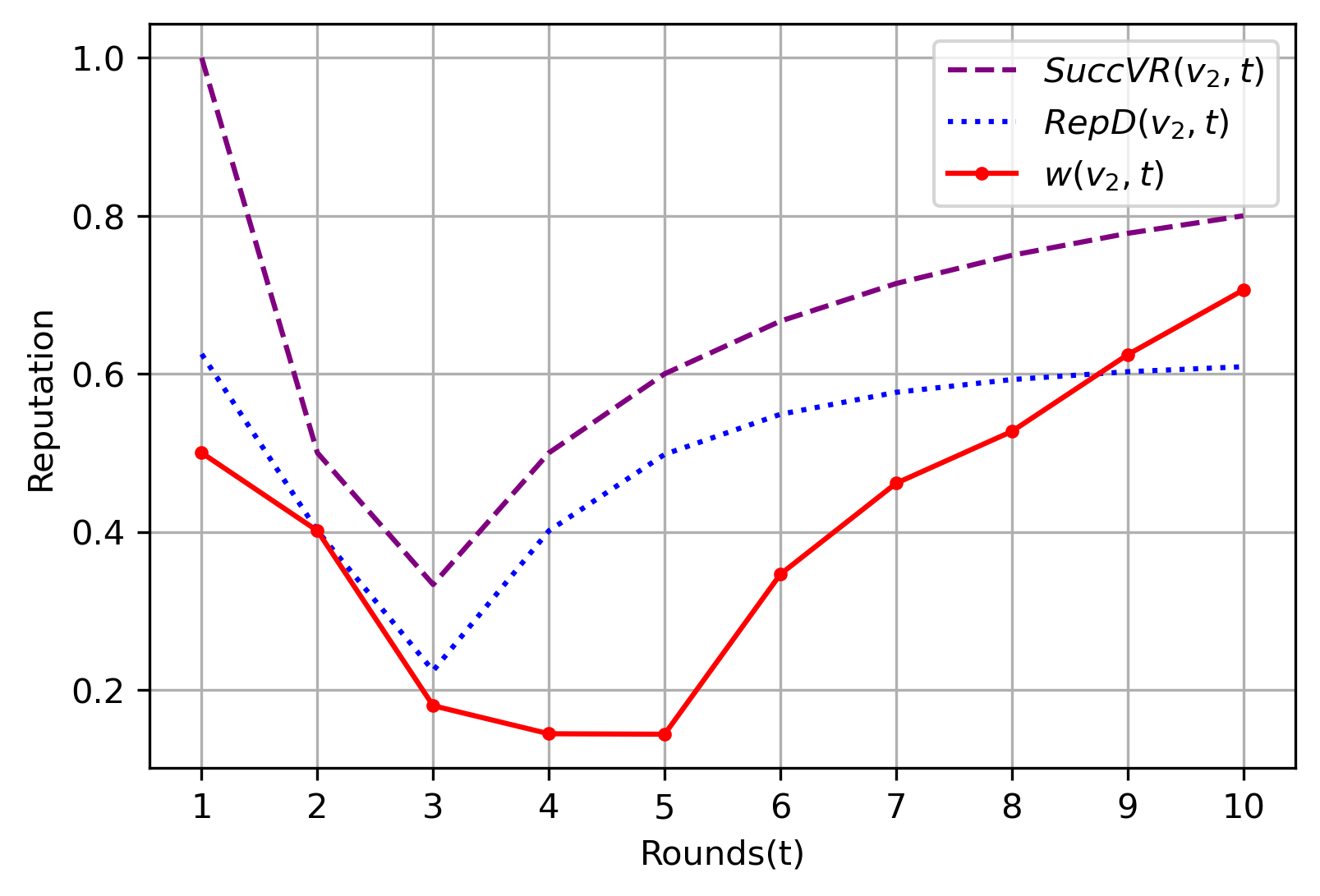}
\caption{Figure plots the results obtained in Table \ref{Tab:main} for the validator $v_2$}
\label{Fig:v2}
\end{figure}

For the validator $v_2$, the outcomes of Table \ref{Tab:main} are depicted in Figure \ref{Fig:v2}. The purple graph (dashed line) represents the successful validation rate $SuccVR(v_2,t)$, the blue graph (dotted line) indicates the reputation degree $RepD(v_2,t)$, and the red graph (continuous line) shows the reputation weight $w(v_2,t)$. These graphs illustrate the reputation behaviour for the validator $v_2$, note that in rounds 2 and 3, the validator was incorrect, therefore, the $SuccVR(v_2,t)$, the $RepD(v_2,t)$, and the $w(v_2,t)$ decrease. For the next rounds, the  $SuccVR(v_2,t)$ and the $RepD(v_2,t)$ increase, nevertheless, $w(v_2,t)$ starts to increase until round 5. 

This experiment shows that the validator $v_1$ made a mistake in the previous round and has the opportunity to correct the mistake in the next round and increase its reputation, as shown in Figure \ref{Fig:v1}. Nevertheless, if the validators were wrong two or more times consecutively, as depicted in Figure \ref{Fig:v2} for the validator $v_2$, the process of increasing and recovering the reputation is slower. This is because repeated mistakes can significantly impact the validator's reputation, leading to a more gradual recovery process. The reputation recovery model is designed to penalise errors more severely, thereby encouraging validators to maintain consistent accuracy and reliability over time. By incorporating such mechanisms, the system ensures that validators who frequently make mistakes face a more challenging path to regain their status, thus promoting higher standards of performance across the network. In the next section, we have analysed the performance of the proposed methodology considering different values for the $\alpha$ parameter.

\subsection{$\alpha $ parameter}\label{Sec:alphaparameter}

With respect to the parameter $\alpha$ introduced in Definition \ref{Eq:repd}, we conducted experiments to analyse its behaviour. These experiments aim to evaluate the performance of the proposed methodology by examining variations in $\alpha$ across ten different values. To this end, we computed the reputation degree $RepD(v_1, t)$ and the reputation weight $w(v_1, t)$ for validator $v_1$, as well as $RepD(v_2, t)$ and $w(v_2, t)$ for validator $v_2$. Tables \ref{Tab:extenv1} and \ref{Tab:extenv2} present the results for $v_1$ and $v_2$, respectively. The first column reports the values of $\alpha$, followed by the corresponding results obtained over twenty rounds.

\begin{table*}[h]
\centering
\begin{adjustbox}{max width=\textwidth}
\begin{tabular}{ccccccccccccccccccccccc} 
\hline
Parameter $\alpha$&Round(t) & 1 & 2 & 3 & 4 & 5& 6&7&8&9&10&11&12&13&14&15&16&17&18&19&20\\\hline
&$SuccVR(v_1, t)$ & 1.00&   0.50&  0.67& 0.75& 0.60&  0.67 &0.71 &0.75 &0.78& 0.80&  0.73 &0.67 &0.69& 0.71&
 0.73& 0.75 &0.71& 0.67& 0.63& 0.65\\
$\alpha=0.1$&$RepD(v_1, t)$ &0.62& 0.58& 0.79& 0.86& 0.72& 0.79& 0.84& 0.86& 0.88& 0.89& 0.85 &0.79&0.82& 0.84&
 0.85& 0.86& 0.83& 0.79& 0.75& 0.78\\
&$w(v_1, t)$ & 0.50&  0.58& 0.82& 0.95& 0.97& 0.99& 1.00&   1.00&   1.00&   1.00&   1.00&   1.00&  1.00&   1.00&
 1.00&   1.00&   1.00&   1.00&   1.00&   1.00 \\\hline
&$SuccVR(v_1, t)$ & 1.00&   0.50&  0.67& 0.75& 0.60&  0.67 &0.71 &0.75 &0.78& 0.80&  0.73 &0.67 &0.69& 0.71&
 0.73& 0.75 &0.71& 0.67& 0.63& 0.65\\
$\alpha=0.5$&$RepD(v_1, t)$ & 0.62& 0.50&  0.68& 0.73& 0.62& 0.68& 0.72& 0.73& 0.74& 0.75& 0.72& 0.68& 0.70&  0.72&
 0.73 &0.73& 0.71& 0.68 &0.65& 0.67\\
&$w(v_1, t)$ & 0.50&  0.50&  0.68& 0.83& 0.87& 0.92& 0.95& 0.98& 0.99& 0.99& 1.00&   1.00&    1.00&   1.00& 
 1.00&    1.00&    1.00&    1.00&    1.00&    1.00\\ \hline
&$SuccVR(v_1, t)$ & 1.00&   0.50&  0.67& 0.75& 0.60&  0.67 &0.71 &0.75 &0.78& 0.80&  0.73 &0.67 &0.69& 0.71&
 0.73& 0.75 &0.71& 0.67& 0.63& 0.65\\
$\alpha=1.0$&$RepD(v_1, t)$ & 0.62& 0.45& 0.61& 0.65& 0.56& 0.61& 0.64& 0.65& 0.66& 0.66& 0.64& 0.61& 0.63& 0.64&
 0.65& 0.65& 0.63& 0.61& 0.58& 0.60\\
&$w(v_1, t)$ & 0.50&  0.45& 0.53 &0.67& 0.71& 0.77& 0.84& 0.89& 0.92& 0.95& 0.96& 0.97& 0.98& 0.98&
 0.99& 0.99& 0.99& 1.00&   1.00&   1.00\\ \hline
&$SuccVR(v_1, t)$ & 1.00&   0.50&  0.67& 0.75& 0.60&  0.67 &0.71 &0.75 &0.78& 0.80&  0.73 &0.67 &0.69& 0.71&
 0.73& 0.75 &0.71& 0.67& 0.63& 0.65\\
$\alpha=1.5$&$RepD(v_1, t)$ & 0.62& 0.42& 0.57& 0.61& 0.52& 0.57& 0.60&  0.61& 0.62& 0.63& 0.60&  0.57& 0.59& 0.60&
 0.61& 0.61& 0.59 &0.57& 0.55& 0.56\\
&$w(v_1, t)$ & 0.50 & 0.42& 0.49& 0.55 &0.57& 0.63& 0.70&  0.77& 0.83& 0.87& 0.90&  0.91& 0.93 &0.94&
 0.95& 0.96& 0.97& 0.97& 0.98 &0.98\\ \hline
&$SuccVR(v_1, t)$ & 1.00&   0.50&  0.67& 0.75& 0.60&  0.67 &0.71 &0.75 &0.78& 0.80&  0.73 &0.67 &0.69& 0.71&
 0.73& 0.75 &0.71& 0.67& 0.63& 0.65\\
$\alpha=2.0$&$RepD(v_1, t)$ & 0.62& 0.40&  0.55& 0.59& 0.50&  0.55& 0.58& 0.59 &0.6&  0.61& 0.58& 0.55& 0.56& 0.58&
 0.59 &0.59& 0.57& 0.55 &0.52& 0.54\\
&$w(v_1, t)$ & 0.50&  0.40&  0.48& 0.53 &0.52& 0.56 &0.63& 0.70&  0.76 &0.81& 0.84& 0.86& 0.88& 0.90&
 0.91& 0.93& 0.94& 0.95& 0.95& 0.95\\\hline
&$SuccVR(v_1, t)$ &1.00&   0.50&  0.67& 0.75& 0.60&  0.67 &0.71 &0.75 &0.78& 0.80&  0.73 &0.67 &0.69& 0.71&
 0.73& 0.75 &0.71& 0.67& 0.63& 0.65\\
$\alpha=2.5$&$RepD(v_1, t)$ & 0.62& 0.39& 0.54& 0.58 &0.49& 0.54& 0.57 &0.58 &0.59 &0.60&  0.57& 0.54& 0.55& 0.57&
 0.57 &0.58& 0.56& 0.54& 0.51& 0.53\\
&$w(v_1, t)$ & 0.50&  0.39& 0.46& 0.52& 0.50&  0.54& 0.60&  0.67& 0.73 &0.78& 0.81 &0.83& 0.85& 0.87&
 0.89 &0.90&  0.92 &0.92& 0.92& 0.93\\ \hline
&$SuccVR(v_1, t)$ & 1.00&   0.50&  0.67& 0.75& 0.60&  0.67 &0.71 &0.75 &0.78& 0.80&  0.73 &0.67 &0.69& 0.71&
 0.73& 0.75 &0.71& 0.67& 0.63& 0.65\\
$\alpha=3.0$&$RepD(v_1, t)$ & 0.62& 0.39& 0.53& 0.58& 0.48 &0.53& 0.56& 0.58& 0.59& 0.60&  0.57& 0.53 &0.55 &0.56&
 0.57& 0.58& 0.55 &0.53& 0.5 & 0.52\\
&$w(v_1, t)$ & 0.50&  0.39 &0.46& 0.52& 0.50&  0.51& 0.57& 0.64 &0.7 & 0.76& 0.79& 0.8&  0.82 &0.84&
 0.86& 0.88 &0.90&  0.90&  0.90&  0.91\\ \hline
&$SuccVR(v_1, t)$ & 1.00&   0.50&  0.67& 0.75& 0.60&  0.67 &0.71 &0.75 &0.78& 0.80&  0.73 &0.67 &0.69& 0.71&
 0.73& 0.75 &0.71& 0.67& 0.63& 0.65\\
$\alpha=3.5$&$RepD(v_1, t)$ & 0.62 &0.38& 0.53& 0.57 &0.47& 0.53& 0.56& 0.57& 0.59& 0.59& 0.56 &0.53 &0.54 &0.56&
 0.57& 0.57& 0.55& 0.53& 0.5&  0.51\\
&$w(v_1, t)$ & 0.50&  0.38 &0.45& 0.51& 0.49& 0.51 &0.56& 0.63 &0.69& 0.75& 0.78& 0.79& 0.81& 0.83&
 0.85& 0.87& 0.89& 0.89 &0.89& 0.90\\ \hline
&$SuccVR(v_1, t)$ & 1.00&   0.50&  0.67& 0.75& 0.60&  0.67 &0.71 &0.75 &0.78& 0.80&  0.73 &0.67 &0.69& 0.71&
 0.73& 0.75 &0.71& 0.67& 0.63& 0.65\\
$\alpha=4.0$&$RepD(v_1, t)$ & 0.62& 0.38& 0.52& 0.57& 0.47& 0.52& 0.55& 0.57& 0.58& 0.59& 0.56& 0.52& 0.54& 0.55&
 0.56& 0.57& 0.55& 0.52& 0.50&  0.50\\
&$w(v_1, t)$ & 0.50 & 0.38 &0.45& 0.51& 0.49& 0.51& 0.56 &0.62 &0.69 &0.75& 0.78 &0.79& 0.80&  0.82&
 0.85& 0.87& 0.88 &0.89& 0.69& 0.70\\ \hline
&$SuccVR(v_1, t)$ & 1.00&   0.50&  0.67& 0.75& 0.60&  0.67 &0.71 &0.75 &0.78& 0.80&  0.73 &0.67 &0.69& 0.71&
 0.73& 0.75 &0.71& 0.67& 0.63& 0.65\\
$\alpha=4.5$&$RepD(v_1, t)$ & 0.62& 0.38& 0.52& 0.57& 0.47& 0.52& 0.55& 0.57& 0.58& 0.59& 0.56& 0.52& 0.54& 0.55&
 0.56& 0.57& 0.55& 0.52& 0.5&  0.51\\
&$w(v_1, t)$ & 0.50&  0.38& 0.45& 0.51& 0.49& 0.51& 0.56& 0.62& 0.68& 0.74& 0.77& 0.78& 0.8&  0.82&
 0.84& 0.87& 0.88& 0.88& 0.69& 0.70\\ 
\hline
\end{tabular}
\end{adjustbox}
\caption{Table shows the successful validation rate $SucVr(v_1, t)$, the reputation degree $RepD(v_1,t)$, and the reputation weight $w(v_1,t)$ for the validator $v_1$ at 20 rounds and multiples $\alpha$ values.}\label{Tab:extenv1}
\end{table*}

\begin{figure*}[!t]
\centering
\includegraphics[width=13.7cm]{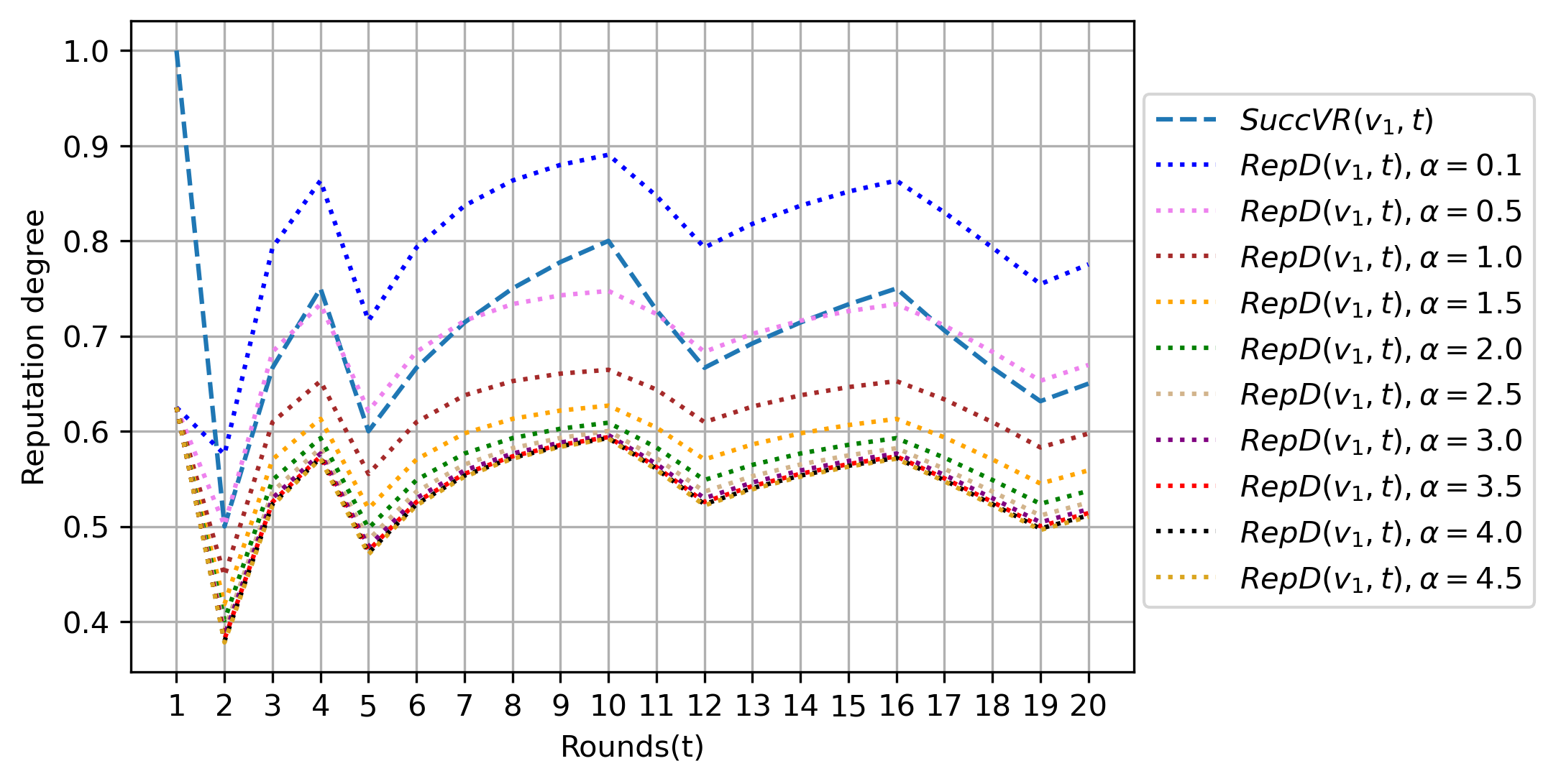}
\caption{Figure shows the plots related to the reputation degree $RepD(v_1,t)$ for the validator $v_1$ as the parameter $\alpha$ varies.}
\label{Fig:repdv1}
\end{figure*}

\begin{figure*}[!t]
\centering
\includegraphics[width=13.7cm]{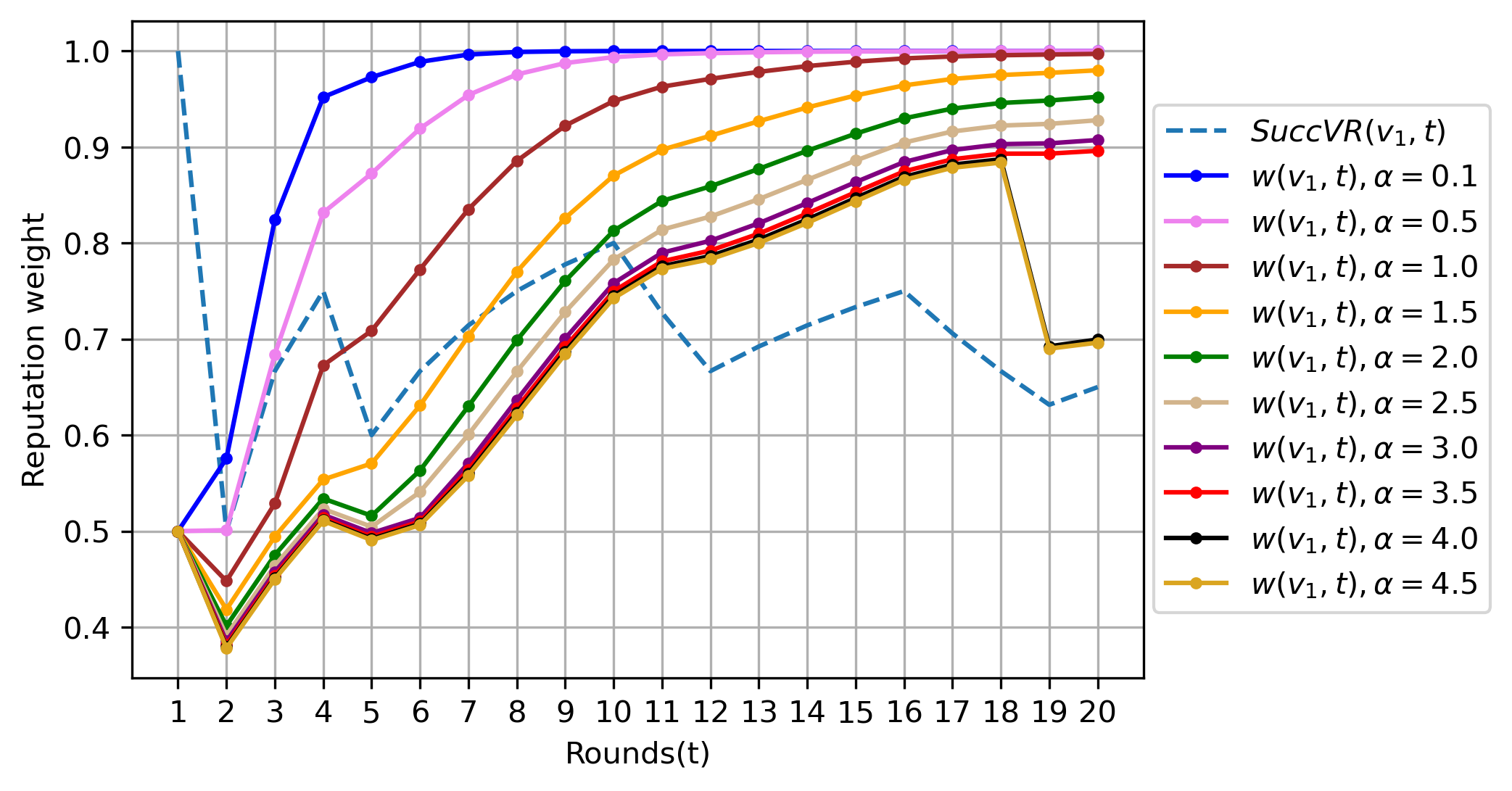}
\caption{Figure shows the plots related to the reputation weight $w(v_1,t)$ for the validator $v_1$ as the parameter $\alpha$ varies. This reputation weight corresponds to the reputation degree illustrated in Figure \ref{Fig:repdv1}.}
\label{Fig:alphasv1}
\end{figure*}

The results presented in Table \ref{Tab:extenv1} are illustrated in Figures \ref{Fig:repdv1} and \ref{Fig:alphasv1}. Figure \ref{Fig:repdv1} displays multiple graphs corresponding to the reputation degree $RepD(v_1, t)$ for different values of $\alpha$. It can be observed that when $\alpha$ is close to zero, the reputation degree assumes higher values. As $\alpha$ increases, the reputation degree decreases accordingly. Figure \ref{Fig:alphasv1} presents the corresponding plots of the reputation weight $w(v_1, t)$, computed using the same values of $\alpha$. These results indicate that the reputation weight is more lenient for small values of $\alpha$, whereas larger values of $\alpha$ lead to more stringent weighting.

\begin{figure*}[!t]
\centering
\includegraphics[width=13.7cm]{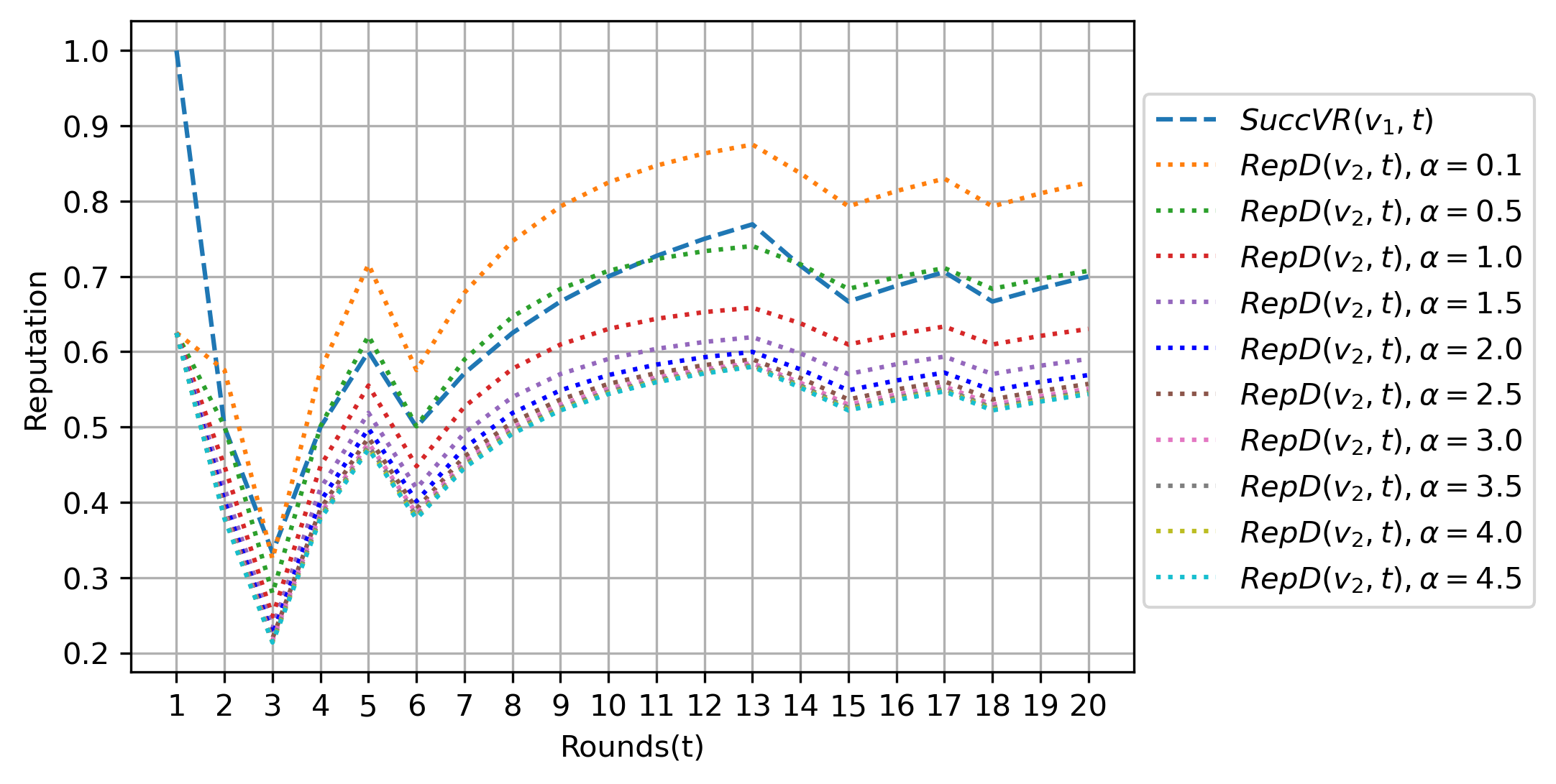}
\caption{Figure shows the plots related to the reputation degree $RepD(v_2,t)$ for the validator $v_2$ as the parameter $\alpha$ varies.}
\label{Fig:repdv2}
\end{figure*}

\begin{figure*}[!t]
\centering
\includegraphics[width=13.7cm]{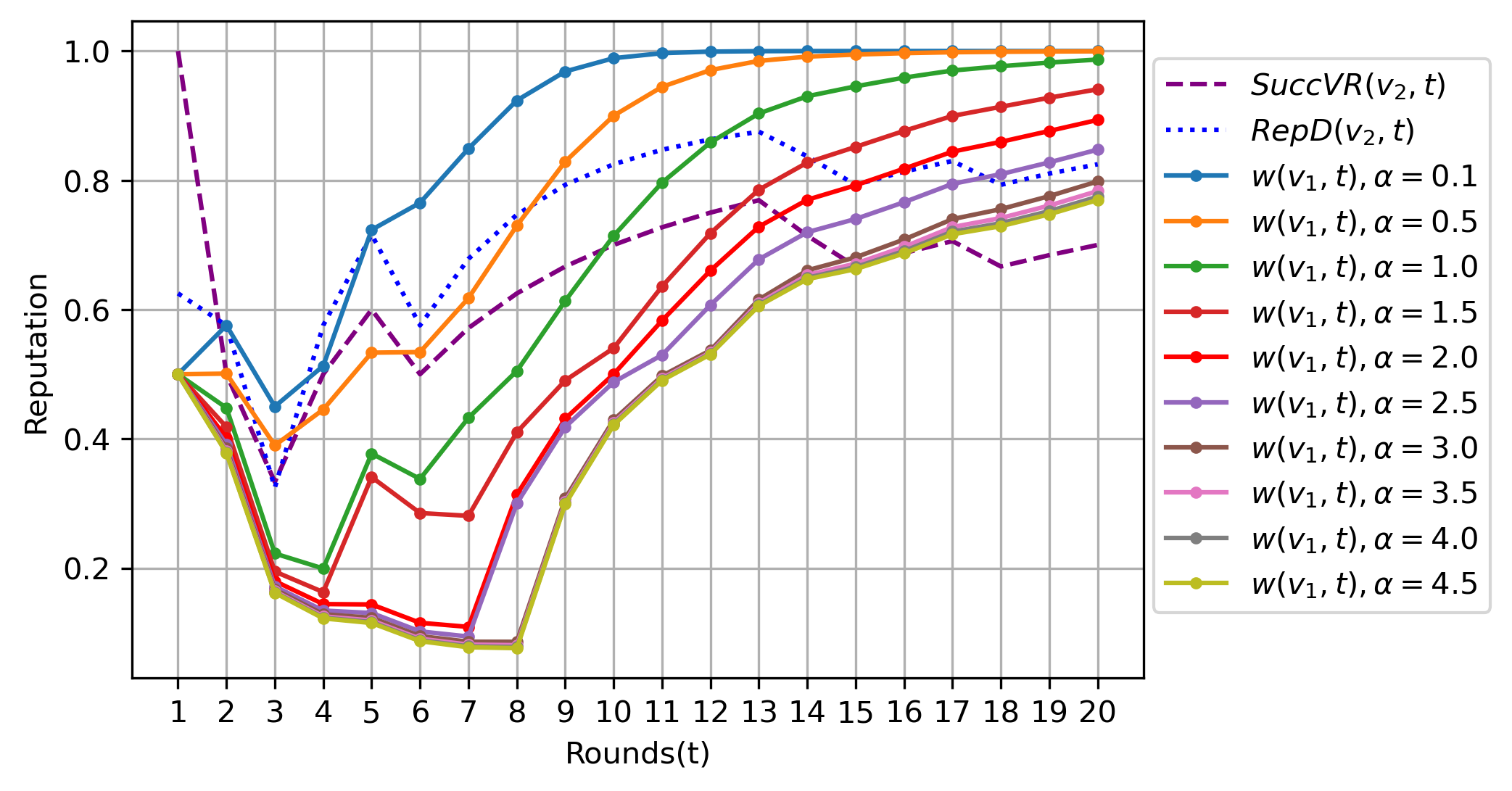}
\caption{Figure shows the plots related to the reputation weight $w(v_2,t)$ for the validator $v_2$ as the parameter $\alpha$ varies.  This reputation weight corresponds to the reputation degree illustrated in Figure \ref{Fig:repdv2}.}
\label{Fig:alphasv2}
\end{figure*}

Similarly, the results in Table \ref{Tab:extenv2} are illustrated in Figures \ref{Fig:repdv2} and \ref{Fig:alphasv2}. Figure \ref{Fig:repdv2} shows the reputation degree $RepD(v_2, t)$ for various $\alpha$ values. When $\alpha$ is close to zero, the reputation degree is higher, and it decreases as $\alpha$ increases. Figure \ref{Fig:alphasv2} displays the reputation weight $w(v_2, t)$ using the same $\alpha$ values. The plots indicate that the reputation weight is more lenient for low $\alpha$ values and becomes more stringent as $\alpha$ increases.

\begin{table*}[!t]
\centering
\begin{adjustbox}{max width=\textwidth}
\begin{tabular}{ccccccccccccccccccccccc} 
\hline
Parameter $\alpha$&Round(t) & 1 & 2 & 3 & 4 & 5& 6&7&8&9&10&11&12&13&14&15&16&17&18&19&20\\\hline
&$SuccVR(v_2, t)$ & 1.00&   0.50&  0.33& 0.50&  0.60&  0.50&  0.57& 0.62& 0.67& 0.70&  0.73& 0.75& 0.77& 0.71&
 0.67& 0.69& 0.71& 0.67& 0.68& 0.70\\
$\alpha=0.1$&$RepD(v_2, t)$ &0.62& 0.58& 0.32& 0.58& 0.72& 0.58& 0.68& 0.75& 0.79 &0.82& 0.85& 0.86& 0.88& 0.84&
 0.79 &0.81& 0.83& 0.79& 0.81& 0.82\\
&$w(v_2, t)$ & 0.50&  0.58& 0.45& 0.51& 0.72& 0.77& 0.85& 0.92& 0.97& 0.99& 1.00&  1.00&   1.00&   1.00& 
 1.00&    1.00&    1.00&   1.00&   1.00&    1.00   \\\hline
&$SuccVR(v_2, t)$ &1.00&   0.50&  0.33& 0.50&  0.60&  0.50&  0.57& 0.62& 0.67& 0.70&  0.73& 0.75& 0.77& 0.71&
 0.67& 0.69& 0.71& 0.67& 0.68& 0.70\\
$\alpha=0.5$&$RepD(v_2, t)$ & 0.62& 0.50&  0.28& 0.50&  0.62& 0.50&  0.59& 0.65& 0.68& 0.71& 0.72& 0.73& 0.74& 0.72&
 0.68 &0.70&  0.71& 0.68& 0.70&  0.71\\
&$w(v_2, t)$ & 0.50 & 0.50&  0.39& 0.45& 0.53& 0.53& 0.62& 0.73& 0.83& 0.90&  0.94& 0.97& 0.98& 0.99&
 0.99&    1.00&    1.00&    1.00&    1.00&    1.00\\ \hline
&$SuccVR(v_1, t)$ &1.00&   0.50&  0.33& 0.50&  0.60&  0.50&  0.57& 0.62& 0.67& 0.70&  0.73& 0.75& 0.77& 0.71&
 0.67& 0.69& 0.71& 0.67& 0.68& 0.70\\
$\alpha=1.0$&$RepD(v_1, t)$ & 0.62& 0.45& 0.25& 0.45& 0.56& 0.45& 0.53& 0.58& 0.61& 0.63& 0.64& 0.65& 0.66& 0.64&
 0.61& 0.62& 0.63& 0.61& 0.62& 0.63\\
&$w(v_1, t)$ & 0.50&  0.45& 0.22& 0.20&  0.38& 0.34& 0.43& 0.51& 0.61& 0.71& 0.80&  0.86& 0.90&  0.93&
 0.95& 0.96& 0.97& 0.98& 0.98& 0.99\\ \hline
&$SuccVR(v_1, t)$ &1.00&   0.50&  0.33& 0.50&  0.60&  0.50&  0.57& 0.62& 0.67& 0.70&  0.73& 0.75& 0.77& 0.71&
 0.67& 0.69& 0.71& 0.67& 0.68& 0.70\\
$\alpha=1.5$&$RepD(v_1, t)$ & 0.62& 0.42& 0.23& 0.42& 0.52& 0.42& 0.49& 0.54 &0.57& 0.59& 0.60&  0.61& 0.62& 0.60&
 0.57& 0.58 &0.59& 0.57& 0.58& 0.59\\
&$w(v_1, t)$ &0.50&  0.42 &0.19& 0.16& 0.34& 0.29& 0.28& 0.41& 0.49& 0.54& 0.64& 0.72& 0.79& 0.83&
 0.85& 0.88& 0.90&  0.91& 0.93& 0.94\\ \hline
&$SuccVR(v_1, t)$ &1.00&   0.50&  0.33& 0.50&  0.60&  0.50&  0.57& 0.62& 0.67& 0.70&  0.73& 0.75& 0.77& 0.71&
 0.67& 0.69& 0.71& 0.67& 0.68& 0.70\\
$\alpha=2.0$&$RepD(v_1, t)$ & 0.62& 0.40&  0.22& 0.40&  0.50&  0.40&  0.47& 0.52& 0.55& 0.57& 0.58& 0.59& 0.60&  0.58&
 0.55 &0.56& 0.57& 0.55& 0.56& 0.57 \\
&$w(v_1, t)$ & 0.50&  0.40&  0.18& 0.14& 0.14 &0.12& 0.11& 0.31& 0.43& 0.50&  0.58& 0.66 &0.73& 0.77&
 0.79& 0.82& 0.84 &0.86& 0.88& 0.89 \\\hline
&$SuccVR(v_1, t)$ &1.00&   0.50&  0.33& 0.50&  0.60&  0.50&  0.57& 0.62& 0.67& 0.70&  0.73& 0.75& 0.77& 0.71&
 0.67& 0.69& 0.71& 0.67& 0.68& 0.70\\
$\alpha=2.5$&$RepD(v_1, t)$ & 0.62& 0.39&0.22& 0.39& 0.49& 0.39 &0.46& 0.51 &0.54& 0.56& 0.57& 0.58& 0.59& 0.57&
 0.54& 0.55 &0.56& 0.54& 0.55& 0.56\\
&$w(v_1, t)$ &0.50&  0.39& 0.17& 0.13& 0.13 &0.10&  0.09& 0.30&  0.42& 0.49& 0.53& 0.61& 0.68& 0.72&
 0.74& 0.77 &0.79& 0.81& 0.83& 0.85 \\ \hline
&$SuccVR(v_1, t)$ &1.00&   0.50&  0.33& 0.50&  0.60&  0.50&  0.57& 0.62& 0.67& 0.70&  0.73& 0.75& 0.77& 0.71&
 0.67& 0.69& 0.71& 0.67& 0.68& 0.70\\
$\alpha=3.0$&$RepD(v_1, t)$ & 0.62& 0.39& 0.22& 0.39& 0.48& 0.39& 0.45& 0.5&  0.53& 0.55& 0.57& 0.58& 0.58& 0.56&
 0.53 &0.54& 0.55& 0.53& 0.54& 0.55\\
&$w(v_1, t)$ & 0.50&  0.39& 0.17& 0.13& 0.12& 0.09& 0.09& 0.09& 0.31& 0.43& 0.5&  0.54& 0.62& 0.66&
 0.68& 0.71 &0.74& 0.76& 0.78& 0.80  \\ \hline
&$SuccVR(v_1, t)$ &1.00&   0.50&  0.33& 0.50&  0.60&  0.50&  0.57& 0.62& 0.67& 0.70&  0.73& 0.75& 0.77& 0.71&
 0.67& 0.69& 0.71& 0.67& 0.68& 0.70\\
$\alpha=3.5$&$RepD(v_1, t)$ & 0.62& 0.38& 0.22 &0.38 &0.47& 0.38 &0.45& 0.50&  0.53& 0.55& 0.56& 0.57& 0.58& 0.56&
 0.53& 0.54& 0.55& 0.53& 0.54& 0.55 \\
&$w(v_1, t)$ & 0.50&  0.38& 0.16& 0.13& 0.12& 0.09& 0.08 &0.08& 0.3&  0.43& 0.49& 0.53& 0.61& 0.65&
 0.67& 0.7 & 0.73& 0.74& 0.76& 0.78\\ \hline
&$SuccVR(v_1, t)$ &1.00&   0.50&  0.33& 0.50&  0.60&  0.50&  0.57& 0.62& 0.67& 0.70&  0.73& 0.75& 0.77& 0.71&
 0.67& 0.69& 0.71& 0.67& 0.68& 0.70\\
$\alpha=4.0$&$RepD(v_1, t)$ & 0.62& 0.38& 0.21& 0.38& 0.47& 0.38& 0.45& 0.49& 0.52 &0.54& 0.56 &0.57& 0.58& 0.55&
 0.52& 0.54& 0.55 &0.52& 0.54& 0.54\\
&$w(v_1, t)$ & 0.50 & 0.38& 0.16& 0.12& 0.12& 0.09& 0.08& 0.08& 0.30&  0.42& 0.49& 0.53& 0.61& 0.65&
 0.67& 0.69& 0.72 &0.73& 0.75& 0.77\\ \hline
&$SuccVR(v_1, t)$ &1.00&   0.50&  0.33& 0.50&  0.60&  0.50&  0.57& 0.62& 0.67& 0.70&  0.73& 0.75& 0.77& 0.71&
 0.67& 0.69& 0.71& 0.67& 0.68& 0.70\\
$\alpha=4.5$&$RepD(v_1, t)$ &0.62& 0.38& 0.21& 0.38& 0.47& 0.38& 0.45& 0.49& 0.52& 0.54& 0.56& 0.57& 0.58& 0.55&
 0.52 &0.54& 0.55& 0.52& 0.53 &0.54\\
&$w(v_1, t)$ & 0.50&  0.38 &0.16& 0.12& 0.12& 0.09& 0.08& 0.08& 0.30&  0.42 &0.49& 0.53& 0.61& 0.65&
 0.66& 0.69& 0.72 &0.73& 0.75& 0.77 \\ 
\hline
\end{tabular}
\end{adjustbox}
\caption{Table shows the successful validation rate $SucVr(v_2, t)$, the reputation degree $RepD(v_2,t)$, and the reputation weight $w(v_2,t)$ for the validator $v_2$ at 20 rounds and multiples $\alpha$ values.}\label{Tab:extenv2}
\end{table*}

The results of this experiment reveal a consistent relationship between the parameter $\alpha$ and both the reputation degree $RepD(v_i, t)$ and the reputation weight $w(v_i, t)$ for the validators. When $\alpha$ is close to zero, both the reputation degree and reputation weight are higher, indicating a more lenient evaluation of validator reputation. Conversely, as $\alpha$ increases, both the reputation degree and reputation weight decrease, reflecting a more stringent assessment. This trend is consistent across different validators, as shown in Tables \ref{Tab:extenv1} and \ref{Tab:extenv2}. Therefore, the choice of $\alpha$ is important in determining the strictness of reputation evaluation, directly influencing how reputation is managed and assessed within the blockchain consensus algorithm.

All source files generated during the development and implementation of the reputation methodology are available at the following GitHub repository: https://github.com/RCBruno/ReputationAwareUninormDriven.

\subsection{Computational Complexity Analysis} \label{Sec:complexanalysis}

This section analyses both the computational and communication complexity associated with the computation of the reputation degree and the reputation weight. Regarding the reputation degree, for each validator $v_i$ at round $t$, the successful validation rate $SuccVR(v_i,t)$ requires one division operation, resulting in constant time complexity $\mathcal{O}(1)$. The evaluation of the membership function $\mu_A(x)$ and the non-membership function $\nu_A(x)$ involves a direct function evaluation (e.g., linear, logarithmic, polynomial, or sigmoid), which also requires $\mathcal{O}(1)$ time. The computation of the intuitionistic index $\pi_A(x)=1-\mu_A(x)-\nu_A(x)$ and the calculation of $RepD(v_i,t)$ involve a fixed number of arithmetic operations and one exponentiation. Therefore, the computation of the reputation degree for a single validator has constant complexity $\mathcal{O}(1)$. For a network with $n$ validators, computing the reputation degree for all validators in a round requires $\mathcal{O}(n)$ time.

The reputation weight $w(v_i,t)$ is obtained by applying the uninorm operator $U(RepD(v_i,t), w(v_i,t-1))$. Since the uninorm (including Fodor's uninorm) is defined as a piecewise function involving a constant number of comparisons and arithmetic operations, its evaluation also has constant time complexity $\mathcal{O}(1)$ per validator. Consequently, updating the reputation weight for all validators requires $\mathcal{O}(n)$ time per round.
Combining both phases, the total computational complexity per round remains linear, $\mathcal{O}(n)$. 

The proposed framework does not require additional peer-to-peer message exchanges beyond those already inherent to the underlying consensus protocol. The successful validation rate $SuccVR(v_i,t)$ is computed locally from validation outcomes that are already broadcast as part of the standard block validation process. Specifically, during each round, validators exchange consensus messages (e.g., block proposals, votes, or confirmations) according to the base consensus algorithm. The reputation mechanism passively observes these outcomes and updates local reputation values. Therefore, the reputation degree and reputation weight calculations introduce no additional all-to-all communication.

Assuming the underlying consensus protocol has communication complexity $\mathcal{O}(C(n))$, the integration of the proposed reputation framework preserves this complexity:
\[
\mathcal{O}(C(n)) + \mathcal{O}(n),
\]
where the additional $\mathcal{O}(n)$ term corresponds only to local updates without extra network-wide broadcasts.

Hence, the proposed intuitionistic fuzzy and uninorm-based methodology introduces negligible communication overhead and maintains scalability even in large-scale blockchain networks.


\section{Discussion and Future directions} \label{Sec:discussion}

\begin{figure}[!t]
\centering
\includegraphics[width=9 cm, height=3.3 cm]{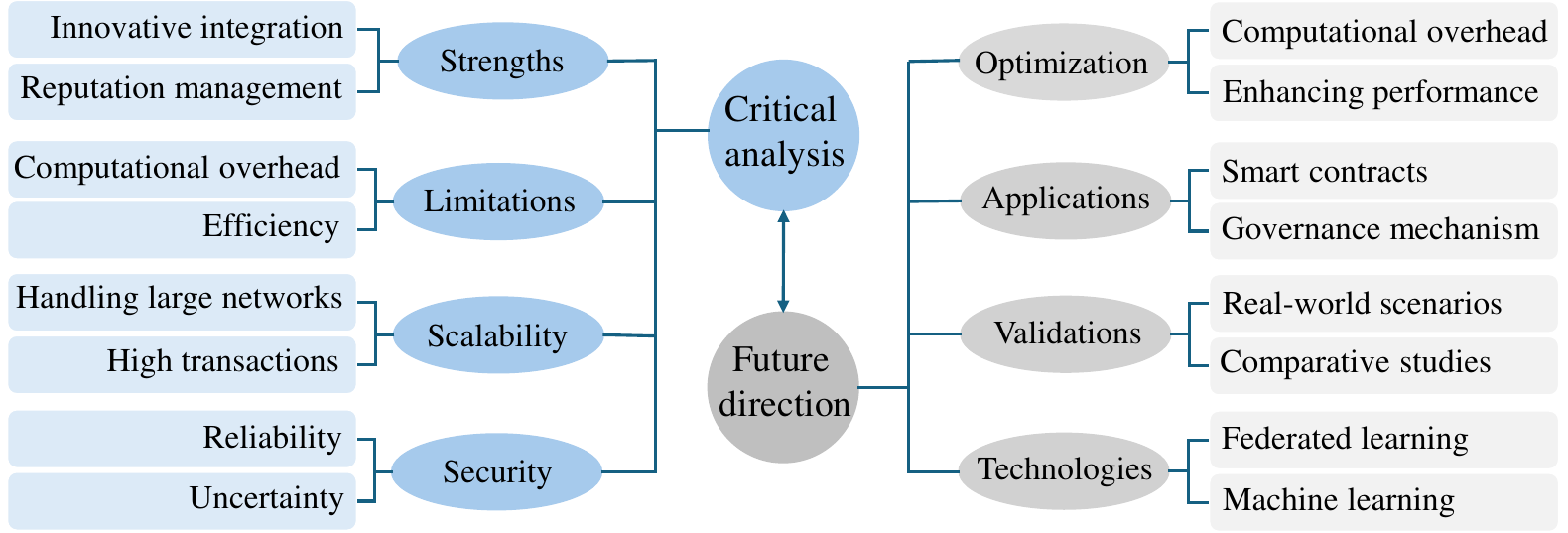}
\caption{Figure summarises the main aspects related to the critical analysis and future directions.}
\label{Fig:critical}
\end{figure}

This work introduces a novel methodology for managing reputation behaviour in consensus algorithms for blockchain, utilising intuitionistic fuzzy sets and uninorm aggregation operations. Figure \ref{Fig:critical} depicts a summary of the main aspects discussed in the critical analysis and future directions sections.

\subsection{Critical analysis}
The integration of intuitionistic fuzzy sets and uninorm aggregation operations offers an innovative solution for managing reputation behaviour in blockchain consensus algorithms, addressing key challenges such as uncertain reputation values and the absence of reputation recovery mechanisms.

The methodological advantage is its ability to handle uncertainty and imprecision in reputation values, enabling validators to recover and improve their reputation even after making errors. This flexibility is demonstrated by the results in Section  \ref{Sec:experiandresults}, which show that validators can regain their standing, promoting a resilient and adaptive network.

Our work serves as a valuable reference for the scientific community, illustrating the potential of integrating IFS and UAO into consensus algorithms where reputation plays a crucial role. In addition, this approach contributes to designing more equitable consensus mechanisms, ensuring diverse and inclusive validator selection processes, thereby mitigating centralisation risks and enhancing network security and fairness.

While the previous paragraphs highlight several strengths of the proposed methodology, we should also consider potential limitations or challenges that have not been addressed. For instance, scalability, further empirical studies and optimisations may be required to ensure that the methodology can handle the exponential growth in transaction volumes and network size typical of large-scale blockchain deployments. For this framework, the overall computational complexity per round is $\mathcal{O}(n)$

Furthermore, integrating IFS and UAO into blockchain consensus algorithms could add computational overhead, which may potentially impact the efficiency and performance of the blockchain. Specifically, the fuzzy set operations and aggregation processes require more sophisticated algorithms and data handling techniques, which might slow down transaction validation processes, particularly in large-scale blockchains.

\subsection{Future directions}
Building on the insights from the critical analysis, we outline potential avenues for future research and development. This includes exploring the scalability of the proposed approach and enhancing its efficiency to ensure optimal performance in large-scale blockchain networks. Extending the application of this methodology beyond consensus algorithms to other areas of blockchain technology, such as smart contract execution and governance mechanisms, could further enhance the system's robustness and adaptability. While the proposed methodology shows promise in addressing key challenges related to reputation management in blockchain consensus algorithms, further exploration and validation are necessary to fully assess its practical implications and scalability in real-world blockchain deployments.

These directions connect to a broader line of fuzzy systems research. Optimising the uninorm parameters, as noted above, could itself be framed as an evolving, self-adaptive fuzzy system that relearns online from streaming validator behaviour rather than being fixed offline, in the tradition of evolving fuzzy classifiers for streaming sensor data~\cite{andreu2010real, andreu2013evolving}. Because reputation is tracked across successive rounds, modelling how a validator's intuitionistic fuzzy reputation value evolves over time is itself a temporal problem, for which time-dependent fuzzy membership functions~\cite{kiani2022temporal} offer a natural formalism. Expressing reputation categories as linguistic, human-interpretable labels rather than raw uninorm outputs, following computing-with-words methodologies~\cite{gupta2022gentle}, and extracting explainable fuzzy rules that justify a given reputation adjustment or validator selection~\cite{fumanal2023artxai}, would make the framework's decisions more auditable to network participants. The large-scale empirical validation called for above would benefit from testing on more realistic, adversarial deployment conditions rather than only simulated scenarios, echoing a similar lesson from clinical brain-computer interface research, where competitions moved from healthy-subject to more challenging patient data to properly stress-test reliability~\cite{chowdhury2021clinical}; the same broader concern, that models must remain reliable under noisy, real-world conditions, extends beyond blockchain consensus to deep learning systems generally~\cite{chaudhary2023review}. Finally, packaging this reputation framework as accessible open-source tooling, in the spirit of toolboxes built for other fuzzy, multivariate analysis pipelines~\cite{andreu2016ealab}, would help practitioners adopt and audit it in deployment.

\section{Conclusion} \label{Sec:conclusion}
In this work, we have explored the areas of intuitionistic fuzzy sets and uninorm aggregation operations to introduce our novel methodology for managing reputation behaviour in consensus algorithms for blockchain. This methodology addresses the challenges associated with uncertain reputation values and a lack of mechanisms for reputation recovery. The results presented in Section \ref{Sec:experiandresults} support the proposed methodology, showing that the validator has the opportunity to increase its reputation even when it was wrong in the previous round, which is not possible concerning other consensus algorithms. Moreover, as Section \ref{Sec:complexanalysis} discussed the framework preserves linear computational complexity per round and does not introduce additional communication overhead beyond that required by the underlying consensus protocol, thereby maintaining scalability for large blockchain networks. This paper provides a reference for the scientific community and could be used in different consensus algorithms where reputation is considered a key parameter. Furthermore, this methodology can be utilised to design equitable consensus algorithms and provide diversification in the selection process to choose the validator node. Integrating intuitionistic fuzzy sets and uninorm aggregation operators into blockchain offers a flexible framework for synthesising uncertain or imprecise information from multiple sources within blockchain networks. Future research directions include large-scale empirical validation, optimisation of the uninorm parameters, and integration with hybrid consensus architectures to further enhance robustness and practical deployment.

\clearpage
\bibliographystyle{elsarticle-num}
\bibliography{references}

\end{document}